\documentclass[10pt,journal]{IEEEtran}			

\usepackage{textcomp}
\usepackage[pdftex]{graphicx}
\graphicspath{{../}}
\DeclareGraphicsExtensions{.pdf,.jpeg,.png,.eps}
\usepackage{fixltx2e}
\usepackage{epstopdf}
\usepackage{cite}
\usepackage{caption}
\usepackage{subcaption}
\usepackage[cmex10]{amsmath}
\usepackage{float,multirow}
\usepackage[table,xcdraw]{xcolor}
\usepackage{amsfonts}
\usepackage[none]{hyphenat}
\usepackage{url}

\usepackage{breakurl}
\usepackage[breaklinks]{hyperref}
\usepackage{soul}
\usepackage{dirtytalk}
\usepackage{hyperref}
\usepackage{comment}
\usepackage[euler]{textgreek}
\usepackage{epsfig,bm,amsmath,amssymb,graphicx,setspace,amsfonts}
\usepackage{algorithm}
\usepackage{color}
\usepackage{xcolor}
\usepackage[T1]{fontenc}
\usepackage{algpseudocode}
\usepackage{textgreek}
\usepackage{amsthm}
\usepackage{array}
\usepackage{balance}
\usepackage{enumerate}
\usepackage{amssymb}
\usepackage{pifont}
\newcommand{\cmark}{\ding{51}}%
\newcommand{\xmark}{\ding{55}}%

\def\BibTeX{{\rm B\kern-.05em{\sc i\kern-.025em b}\kern-.08em
        T\kern-.1667em\lower.7ex\hbox{E}\kern-.125emX}}
\begin{document}

\title{Fast, Reliable, and Secure Drone Communication: A Comprehensive Survey}

			 \author{Vikas Hassija, Vinay Chamola, \textit{Senior Member, IEEE}, Adhar Agrawal, Adit Goyal, Nguyen Cong Luong, Dusit Niyato, \textit{Fellow, IEEE}, F. Richard Yu, \textit{Fellow, IEEE}, and Mohsen Guizani, \textit{Fellow, IEEE}
			 \thanks{Manuscript received July 11, 2020, revised December 25th, 2020 and April 26th, 2021 Accepted XXXX (IEEE Communications Surveys and Tutorials)}
  \thanks{Vikas Hassija, Adhar Agrawal, and Adit Goyal are with the Department of Computer Science and IT, Jaypee Institute of Information Technology, Noida, India 201304 (e-mail: vikas.hassija@jiit.ac.in, adharagrawal98@gmail.com, aditgoyal@hotmail.com).}%
		\thanks{Vinay Chamola is with the Department of Electrical and Electronics Engineering \& APPCAIR, BITS-Pilani, Pilani Campus, India 333031 (e-mail: vinay.chamola@pilani.bits-pilani.ac.in).}%
		\thanks{ N. C. Luong is with the Faculty of Information Technology, PHENIKAA University, Hanoi 12116, Vietnam, and is with PHENIKAA Research and Technology Institute (PRATI), A\&A Green Phoenix Group JSC, No. 167 Hoang Ngan, Trung Hoa, Cau Giay, Hanoi 11313, Vietnam (email: luong.nguyencong@phenikaa-uni.edu.vn).}
		\thanks{ Dusit Niyato is with School of Computer Science and Engineering, Nanyang Technological University, Singapore (e-mail: dniyato@ntu.edu.sg).}
		\thanks{ F. Richard Yu, \textit{Fellow, IEEE} is with the Department of Systems and Computer
Engineering, Carleton University, Ottawa, ON K1S 5B6, Canada (e-mail: richard\_yu@carleton.ca).}
\thanks{Mohsen Guizani is with Computer Science and Engineering Department, Qatar University, Qatar (e-mail: mguizani@ieee.org).}
 \thanks{Digital Object Identifier: XXXXXXXXXXXX}}

	\maketitle
	
\begin{abstract}
Drone security is currently a major topic of discussion among researchers and industrialists. Although there are multiple applications of drones, if the security challenges are not anticipated and required architectural changes are not made, the upcoming drone applications will not be able to serve their actual purpose. Therefore, in this paper, we present a detailed review of the security-critical drone applications, and security-related challenges in drone communication such as DoS attacks, Man-in-the-middle attacks, De-Authentication attacks, and so on. Furthermore, as part of solution architectures, the use of Blockchain, Software Defined Networks (SDN), Machine Learning, and Fog/Edge computing are discussed as these are the most emerging technologies. Drones are highly resource-constrained devices and therefore it is not possible to deploy heavy security algorithms on board. Blockchain can be used to cryptographically store all the data that is sent to/from the drones, thereby saving it from tampering and eavesdropping. Various ML algorithms can be used to detect malicious drones in the network and to detect safe routes. Additionally, the SDN technology can be used to make the drone network reliable by allowing the controller to keep a close check on data traffic, and fog computing can be used to keep the computation capabilities closer to the drones without overloading them. 
\end{abstract}
    
    \IEEEpeerreviewmaketitle


\section{Introduction}
 
\textcolor{black}{The number of drone or UAV based applications are drastically increasing. Based on the latest TechSci report, the overall revenue from the drone application-related market is expected to drastically improve from 69 billion dollars in 2018 to 141 billion dollars in $2023$ \cite{techsci}. The first application of drones was seen in $1849$ when the Austrian army attacked Venice with some unmanned balloons filled with explosives \cite{Austrian}. This was the point where the idea of drones and its related applications came into the picture and became a topic of exploration for researchers. 
During WWII, Reginald Denny invented the first remote-controlled aircraft called the Radioplane OQ-2. It was the first mass-produced UAV product in the US, and was a breakthrough in manufacturing and supplying drones for the military \cite{firstdrone}. The use of drones in multiple domains has been rapidly increasing in the past few years. }
 
  \begin{figure}[!t]
 	\centering
 	\includegraphics[width=90mm]{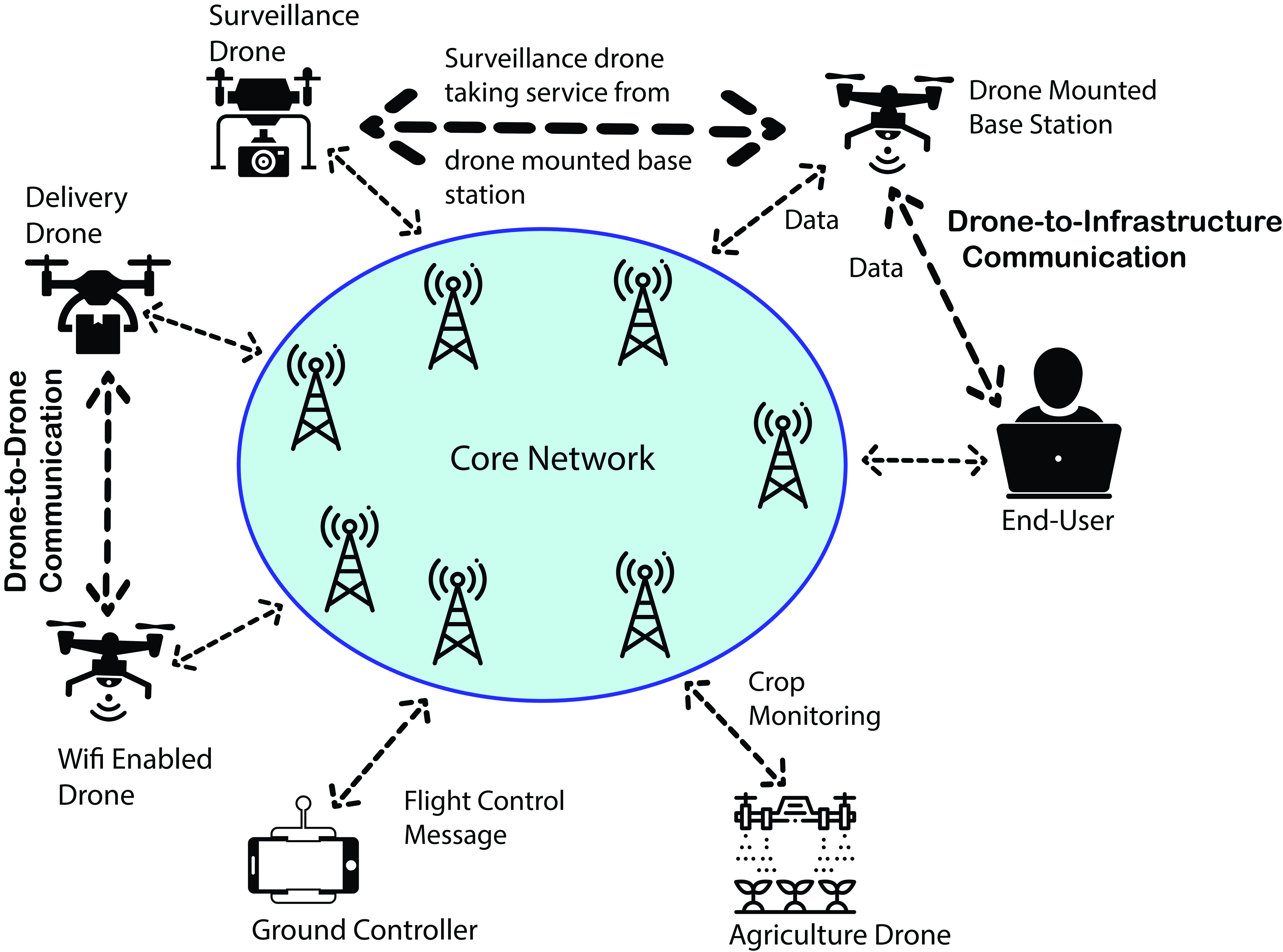}
 	\caption{\textcolor{black}{Basic Process of Drone Communication \cite{fig1}.}}
 	\label{droneWorking}
 \end{figure}
  
\textcolor{black}{Drones work on a simple procedure that involves a data link from the ground controller to the drone and a data link from the drone to the satellite. The ground station controller is also in link with the satellite at every point of time. The basic functioning of drone communication is pictorially shown in Fig. \ref{droneWorking}. Communication between the drone and the other components takes place through radio waves. Drones can help in sending data from one point to another with low latency \cite{fig1}. 
Drones can provide on-the-fly communication facilities in areas where terrestrial infrastructure is poor or has been destroyed, and to provide any further destruction or harm, emergency services are required in disaster-struck areas \cite{drone_ml}. UAVs can act as a communication-bridge between ground users and network nodes. Furthermore, they can also be used in various monitoring or surveillance operations. 
A 3-D network can also be made to integrate drone base stations (droneBS) and cellular-connected drone users \cite{rev2}. Although these applications are highly promising to provide safety and comfort to all, they can also bring disastrous results if the drone communication links are hacked and misused. Being resource-constrained, drones are highly vulnerable to physical and cyber attacks/threats \cite{cyber}. The storage and battery capacity of drones is limited and if proper care is not taken, it is easy to hack the chips and the sensors installed inside the drone's circuit to get all the stored information. Therefore, it is highly imperative to focus on the security standards for drone communication as their applications increase \cite{newly1, newly2}. The authors of \cite{rev1} propose a way to reduce the service-time of drones. 
Drone path planning can be done for secure positioning and it's verification of various components \cite{conti4}.}
 
\begin{figure*}[!t]
 	\centering
 	\includegraphics[width=180mm]{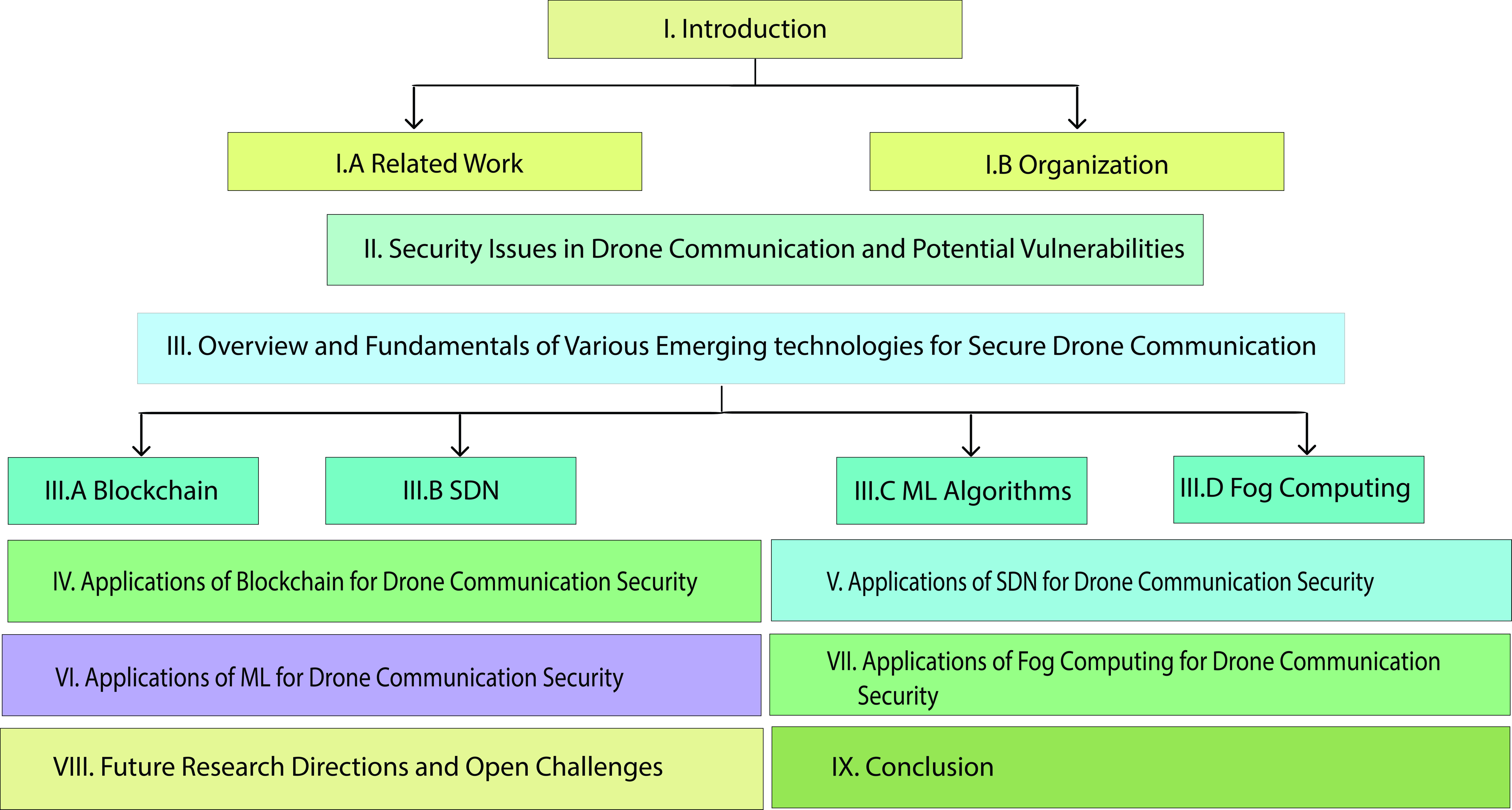}
 	\caption{\textcolor{black}{Structure of our Survey.}}
 	\label{organization}
 \end{figure*}
  
    	\begin{table}[!t]
 	\caption{List of Major Acronyms}
 		\centering
 		\resizebox{1\columnwidth}{!}{
 		\begin{tabular}{|l|l|}
 			\hline \rowcolor[gray]{0.7}
 			\textbf{{Notation}} & \textbf{{Meaning}}\\ 
 			\hline 
         ADMM & Alternating Direction Method of Multipliers \\
         \hline
 			AODV & Ad hoc On demand distance vector\\
 			\hline
         ApaaS & Authentication Proxy as a Service \\
         \hline
            CPS & Cyber-Physical System \\
            \hline
           FANET & Flying Ad-hoc Network  \\
           \hline
           FQ & Fair Queuing \\
           \hline
         \textcolor{black}{GNSS} & \textcolor{black}{Global Navigation Satellite System Signals} \\
         \hline
          HAE & Homomorphic Authenticated Encryption\\
           \hline
          ILP & Integer Linear Program \\
          \hline
       IMU & Inertial Measurement Unit   \\
           \hline
           \textcolor{black}{LOIC} & \textcolor{black}{Low Orbit Ion Cannon} \\
           \hline
     \textcolor{black}{LiDAR} & \textcolor{black}{Light Detection And Ranging} \\
     \hline
      NBTM & Neural-Blockchain Based Transport Model\\
        \hline
     NIDS & Network intrusion detection systems \\
  	    \hline
     PKG & Private Key Generator \\
     \hline
    PUF & Physical Unclonable Function \\
    \hline
     \textcolor{black} {RSA} & \textcolor{black}{Rivest-Shamir-Adleman} \\
      \hline
            TVA+ & Traffic Validation Architecture \\
            \hline
         VIO & Visual Inertial Odometry \\
         \hline
 		\end{tabular}	}
 			\label{Acronymtable}
 	\end{table}
 	
\begin{table*}[!t]
 	\caption{Related Surveys on Drone Security}
 		\centering
 			\resizebox{1\textwidth}{!}{
 		\begin{tabular}{|l|l|l|}
 					\hline \rowcolor[gray]{0.7}
 					\textbf{Year} & \textbf{Author} & \textbf{Contributions} \\
					
 					\hline
 		2015 & Lav Gupta et al.,  \cite{rw3}
  &
 	Discussion on security issues in swarm of drones or Flying Ad-hoc Network (FANET)	\\
 		          \hline
 	2016 & Riham Altawy, Amr M. Youssef , \cite{civilian}
 &
     Survey on security, privacy, and safety aspects of civilian drones   \\    
 		            \hline
 	2016 & Samira Hayat  et al., \cite{rw4}
  & Discussion on requirements of UAV networks for upcoming drone applications
 \\
 				    \hline
     					2017 & Mohammad Mozaffari,  Walid Saad  et al., \cite{rw5}
  & Issues that UAV faces due to wireless networks
 \\
 				    \hline
 	2018 &  Silvia Sekander, Hina Tabassum et al., \cite{rw6}

  & Issues due to wireless networks and the architecture of 5G for UAV \\
                 	\hline
 				2019 & Azade Fotouhi et al., \cite{rw2}
  & \textcolor{black}{Challenges faced by UAV in cellular communication}
   \\
 					\hline
 				2019  & Saeed H. Alsamhi et al., \cite{rw1} & The challenges faced in collaboration of drones and IoT specifically for smart cities
   \\
                     \hline
                 2019 & Sun Xingming, Yueyan Zhi et al., \cite{tab1} & Survey on security and privacy issues of UAV \\
 					\hline
 				 \textcolor{black}{2021} & \textcolor{black}{This paper} & \textcolor{black}{Survey on existing and upcoming security challenges in drone communication and their solutions}\\
 				    \hline
 			\end{tabular}	}
 			\label{table_relatedworks}
 	\end{table*}

\textcolor{black}{Due to the increasing use of drones, the issues related to drone security, privacy, reliability, regulation, and ownership are also increasing at the same pace. 
There are various security-critical applications where drones fail to provide complete security of data, and that results in a great loss and life-threatening risk.  For example, on $29^{th}$, November $2018$, a drone was hacked in Las Vegas and it came into the path of a tour helicopter \cite{2}. Fortunately, the pilot could manage to avoid a crash, but this may not be the case in all such events. A crash might have resulted in the loss of life of many civilians. The incident was investigated by the Federal Aviation Administration (FAA) and some strict rules against drone usage were also brought into action. Various such threats can be caused by the unrestricted use of drones in different applications without any standard security parameters. In this section, we present various important drone applications that are associated with critical security issues. Table \ref{Acronymtable} shows the list of major acronyms used throughout this survey. }

\subsection{Related Surveys and Our Contributions}

Although a few recent works focus on surveys of issues related to drone communications, 
the existing surveys generally consider a specific domain or utility of drones. For example, the authors of \cite{rw1} provide a detailed survey on the challenges faced in the collaboration of drones and IoT specifically for smart cities. Another work presented in \cite{civilian} discusses the security, privacy, and safety aspects specific to civilian drones.

Furthermore, a significant number of surveys have been done earlier for discussing the privacy and security issues present specifically in UAVs or communication networks. The authors of \cite{rw2} focus on the use of UAVs for cellular communications. The authors discuss various standardization advancements, practical aspects, regulatory issues, and security challenges related to the use of UAVs in cellular communication. The authors of \cite{rw3} provide a full review of various security challenges faced in UAV communication networks. The authors specifically focus on the issues faced in a swarm of drones or Flying Ad-hoc Network (FANET). A comparative study of issues that differentiate FANET from other ad-hoc networks such as Mobile Ad-hoc Network (MANET) or Vehicular Ad-hoc Network (VANET) is also done in good detail. Furthermore, the authors of \cite{cong1} review the use of Game Theory-based approaches for UAVs. UAV path deviation attacks have been surveyed in \cite{conti2}. The authors of \cite{rw4} provide a review of the characteristics and requirements of UAV networks for upcoming drone applications. A generic review on all the network-related requirements such as safety, scalability, privacy, connectivity, security, and adaptability are discussed. The authors of \cite{rw5} and \cite{rw6} also emphasize on the issues related to the use of UAVs in the wireless network. The work done in \cite{rw5} provide some key guidelines over analyzing, designing and optimizing communication systems unique to UAVs. The authors also discuss the need for various security measures required for drones. A complete drone architecture for 5G has been presented in good detail. Moreover, a comprehensive survey discussing the security and privacy issues faced by UAVs is presented in \cite{tab1}.

Hence, different from any of the previous works, this work is a comprehensive survey on \textcolor{black}{the most critical} existing and upcoming security challenges in drone communication and the related solutions. This paper aims to help the readers get an overview of the state-of-the-art security challenges in drone communication. The readers will also have a good overview of existing and emerging security solutions for drone communication. Table \ref{table_relatedworks} shows the major survey works done in the direction of drone security in last few years. 
 
The main contributions of this work are as follows:

 \begin{enumerate}
 \item[{\bf 1.}] {A complete review of different existing and anticipated attacks in drone communication.}
 \item[{\bf 2.}] {Detailed and realistic recommendations to improve the drone application architecture for secure communication.}
 \item[{\bf 3.}] {Extensive analysis on the existing and upcoming solutions that empower the use of drone communication in multiple domains.}
 \item[{\bf 4.}] {An assessment of the future research areas, existing challenges, and, open issues for developing secure drone applications.}
    
 \end{enumerate}

\subsection{Organization}
The rest of the paper is organized as follows. In Section \ref{sec2}, we discuss various security issues and security-critical applications of UAVs in different domains. Section \ref{sec3} discusses the fundamentals of various emerging technologies for secure drone communication. Four major drone communication security approaches, i.e., Blockchain, Software Defined Networks (SDN), Machine learning, and Fog/edge computing are presented in Section \ref{sec5}, \ref{sec6}, \ref{sec7}, and \ref{sec8}, respectively. Section \ref{sec9} describes various future research areas, existing challenges, and open issues in drone security. Finally, we conclude the paper in Section \ref{sec10}. The organization of the survey is also shown in Fig. \ref{organization}.

\section{\textcolor{black}{Security Issues in Drone Communication and Potential Vulnerabilities}}
\label{sec2}Drone communication faces some specific security challenges along with the generic cyber-threats. One of the reasons for these specific issues is that drones are unmanned and it is difficult to handle or prevent unanticipated issues dynamically and adaptively. Special attention needs to be given to drone security issues as drones are different from the traditional IoT devices (mobile phones, sensor-based alarms, smart trackers, etc.), and we need drones to adapt to several advanced security concepts, such as confidentiality, authentication, access control, and data protection, while being highly resource constrained devices. Usage of drones needs to take care of vulnerability concerns from sensor networks, mobile communication networks, the internet, et cetera. The drones communicating via cellular data use radio signals to communicate with the controller. The controller sends the radio signals through its transmitter and these are received by the drone through its receiver. The radio signals in between can be jammed or can be tampered \cite{radio}. As stated by an IBM researcher, drones can be hijacked easily if they do not have encryption on their onboard chips \cite{10}. Because of resource constraint issues of drones, encryption would not be the ideal solution. With a huge amount of data exchange in drone communications, encryption and decryption using complex algorithms require a certain amount of computational power. The security concerns become more severe when drones use Wi-Fi for communication \cite{norton}. \textcolor{black}{Table \ref{IoT_diff} summarizes how susceptible various wireless communication networks are in comparison with drone communication systems.} In this section, we present various specific security challenges faced by drones. Furthermore, we also discuss the specific security vulnerabilities for each attack in some drone applications. Various ideas and methodologies to overcome these security challenges are discussed in the upcoming sections of this paper.

\begin{table*}[]
\color{black}
\centering
\caption{Difference between security vulnerabilities of different wireless networks}
\begin{tabular}{|l|l|l|l|l|l|}
\hline \rowcolor[gray]{0.8}
Security Issue & Ad-hoc Network & Sensor Network & Mesh Network & Vehicular Network & Drone Comm.   \\
\hline
DoS & {\cmark} (Low) & {\cmark} (High) & {\cmark} (Low) & {\cmark} (Medium) & {\cmark} (High)\\
\hline
Man-in-the-middle & {\cmark} (Medium) & {\cmark} (High) & {\cmark} (Low) & {\cmark} (Low) & {\cmark} (High)\\
\hline
GPS spoofing & {\xmark} & {\xmark} & {\xmark} & {\cmark} (Medium) & {\cmark} (High)\\
\hline
Radar & {\xmark} & {\xmark} & {\xmark} & {\xmark} & {\cmark} (High)\\
\hline
Jamming & {\cmark} (Medium) & {\xmark} & {\cmark} (Low) & {\cmark} (Low) & {\cmark} (High)\\
\hline
Wormhole & {\cmark} (High) & {\cmark} (High) & {\cmark} (Low) & {\cmark} (Low) & {\cmark} (High)\\
\hline
\end{tabular}
\label{IoT_diff}
\end{table*}
   
\subsection{Security Issues In Drones}
Few of the security threats discussed below are more drone-specific (GPS spoofing, radars, jamming, and wormhole attacks), whereas the relatively generic issues mentioned are discussed based on how adversaries can exploit them to threaten the use of drones.
 
\subsubsection{ Denial of Service Attacks}
\begin{figure}[!t]
	\centering
	\includegraphics[width=90mm]{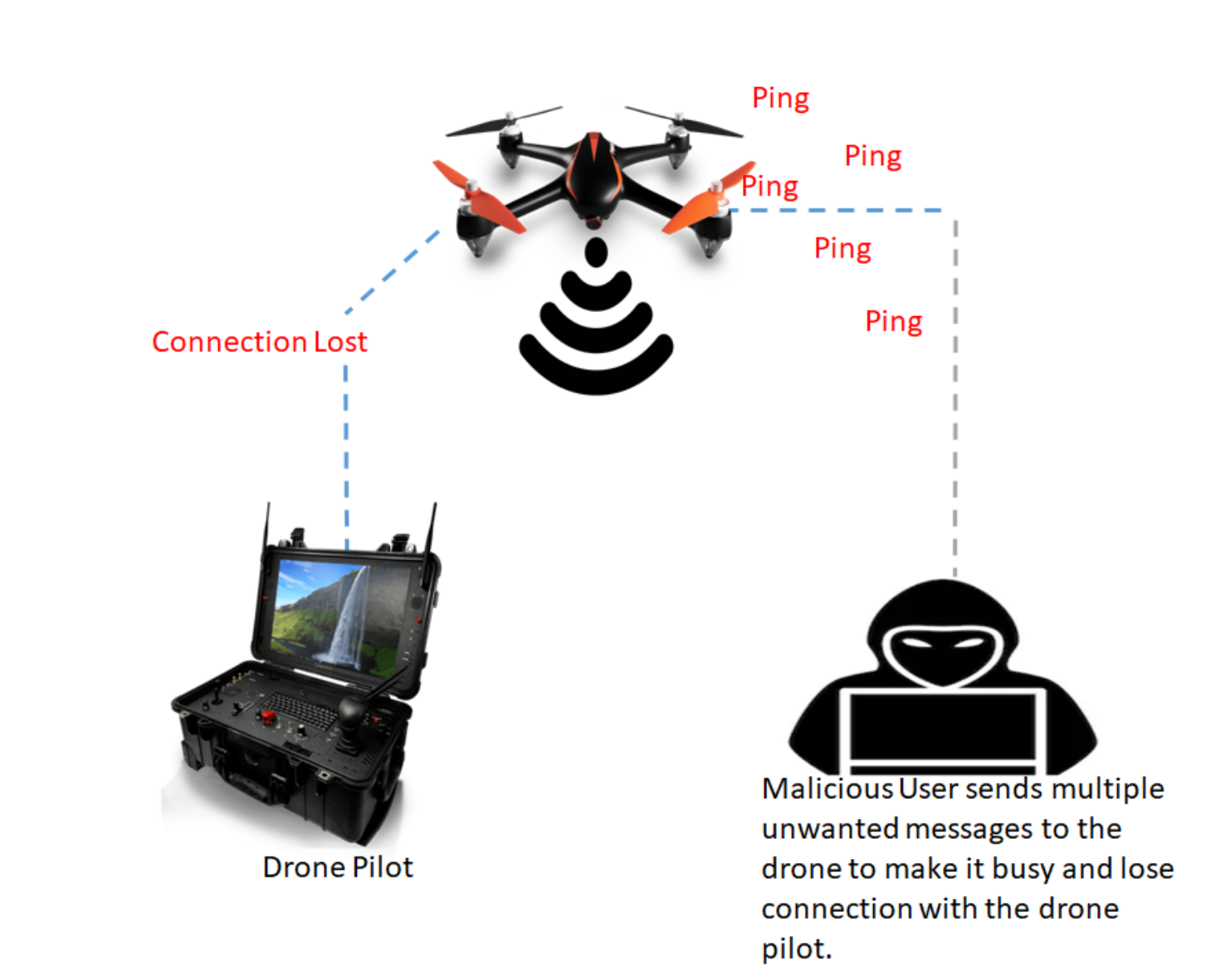}
	\caption{Denial of Service Attack on Wi-Fi Enabled Drone.}
	\label{dos}
\end{figure}

The DoS or the Denial Of Service is the most common and easy type of attack that an adversary can use to stop the drone from functioning normally. This is the simplest way of entering into the drone network and making it useless or sometimes even harmful \cite{newly14}. Fig. \ref{dos} shows the basic working of the DoS attack in case of drone communication. Due to a large number of superfluous requests, the access of shared resources to legitimate users is restricted. This will cause the system to overload, and might result in rejection of some or all legitimate requests to be fulfilled. In this process, the network connection between the ground controller and the drone is de-authenticated as the adversary sends several data packets to the drone which leads to the failure of the computational power of the drone \cite{DoS}. Data packets can be easily created by any packet generator application and can be sent directly to the drone's network. ICMP (Internet Control Message Protocol) packets will be sent at a very high rate which will make the network of the drone overflow, resulting in the loss of control of the drone both by the drone as well as the ground controller. It is also possible that there is some malicious code present in one of the sent data packets, that can be used to attack the drone. Such attacks can be used by the hijacker to crash the drone causing harm to civilians and government agencies.
   
The authors of \cite{doscase} have demonstrated the effects of DoS attack on $2$ types of drones namely, Augmented Reality aerial drone (AR Drone) $2.0$ which is a cheap quadcopter and $3$DR Solo which is a costlier quadcopter. The authors experimentally evaluate three DOS attack-tools: Netwox, Low Orbit Ion Cannon (LOIC), and Hping3, to analyze the drone's behavior. Both these types of drones are widely available in the markets. Both the drones were tested for the image quality transmitted and how DoS attacks affect them. The study found that both the costlier and cheap drones show a significant drop in frame rates demonstrating clearly to us that even premium drone manufacturers are not paying enough attention to drone security. The increase in network latency shows the ease of DoS attacks even on such high end drones.
 
\textit{De-Authentication Attacks:}
This is a type of attack that can make the use of drones difficult in various applications. This is a specific type of Denial Of Service attack in which communication is disrupted between the client and the Wi-Fi access point. In this attack, the control of the drone is lost by the pilot as the attacker de-authenticates the ground pilot. Attacker can send a de-authentication frame to a wireless access point at any point in time as encryption is not needed to send the frame, despite the privacy technique employed \cite{dos_encrypt}. The attacker only needs the mac address of the drone which is made available through any of the tools like `Aircrack-ng' \cite{11}. In the de-authentication attack, the drone is hijacked by using this tool which specifies the mac address of the drone. As soon as the Aircrack-ng tool is activated, the connection between the drone and the ground controller is de-authenticated. The attacker can use this tool to communicate with the drone and direct it maliciously. This attack makes the drone go out of control and leads to a heavy loss.

De-authentication attacks have become one of the newest concerns in the industry as e-commerce giants, such as Amazon, look towards product delivery mechanisms for drones. One of the most famous methods for carrying out this attack is SkyJack \cite{decase} which uses an AR. Drone $2.0$ along with a Raspberry Pi and wireless Adapters to hack and control drones. It sends de-authentication requests through Aircrack-ng which is used to disconnect the target drone from their user and then use the node-ar-drone library to communicate with the target drone.

\begin{figure}[!b]
    \centering
    \includegraphics[width=90mm]{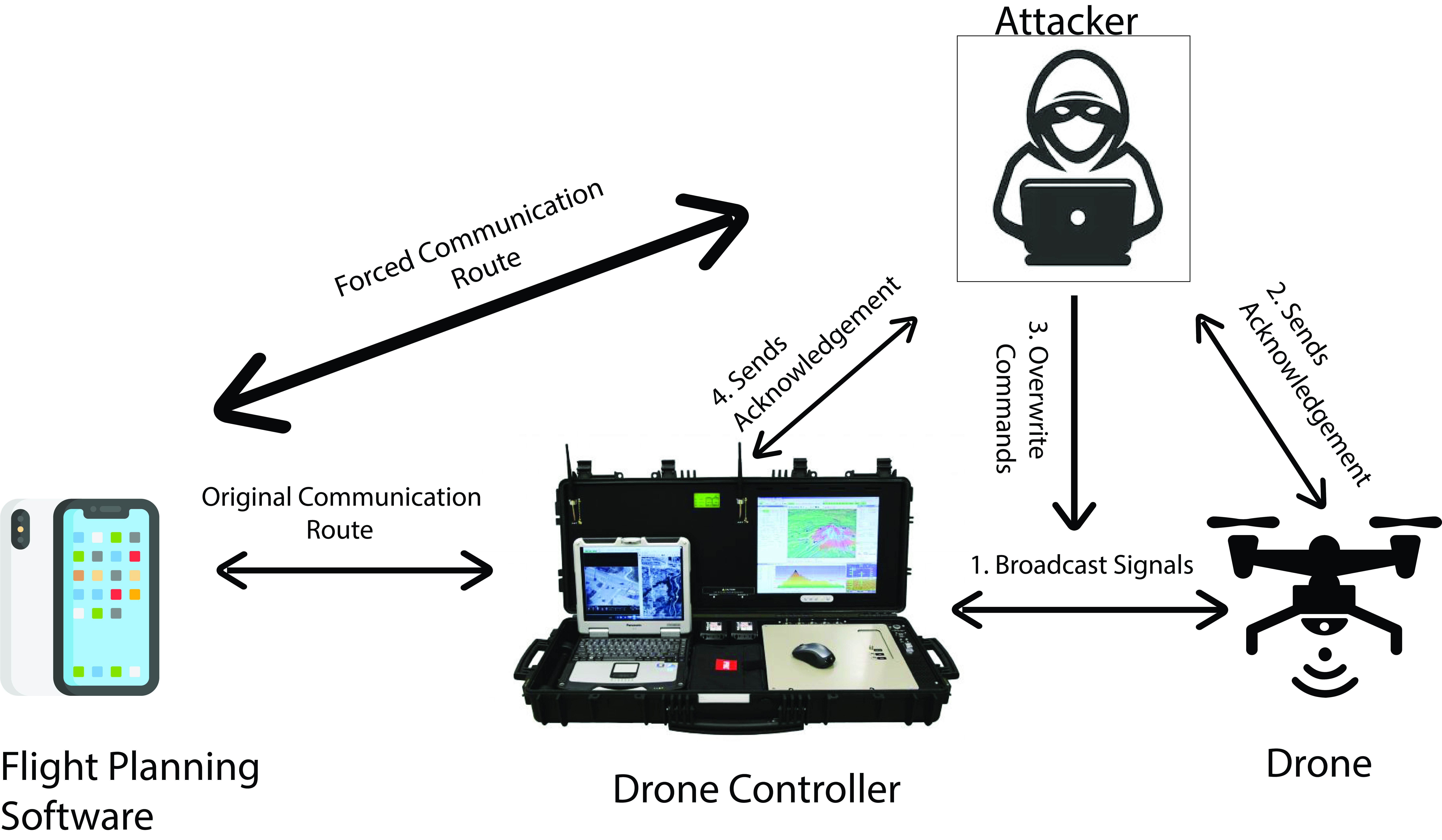}
    \caption{Man-In-The-Middle Attack On Drone \cite{fig4}.}
    \label{manattack}
\end{figure}

\subsubsection{Man-In-The-Middle Attack}
Man-in-the-middle attack places an adversary in between the client and the drone. The adversary uses a device known as Wi-Fi Pineapple \cite{11}. Fig. \ref{manattack} represents the implementation of man-in-the-middle attack. In this attack, the flight planning software broadcasts the plan to the drone controller which sends it further to the drone. On successfully receiving the commands from the controller, the drone sends the acknowledgement, which is received by the attacker in between the drone and the controller. The attacker uses Wi-fi Pineapple to send the forced commands to the controller. Once the pineapple is set, it will run the recon mode which will trace out all the possible access points that the client software may be using. Once the access point (drone) is traced, it is added to the  Pine-AP SSID (Service Set Identifier) pool. This command is forwarded to the drone and the actions intended by the adversary are imperceivably implemented by the drone. One example of man-in-the-middle attack is active eavesdropping \cite{eavs} in which the adversary connects himself with the drone controller. After getting the access of the drone through the SSID of the drone, the hacker sends the fake commands to the controller, making them believe that they are communicating with the drone itself \cite{newly13}.

Fig. \ref{manattack} depicts the man-in-the-middle attack. The authors of \cite{fig4} explored various vulnerabilities of UAVs; if a weak encryption scheme is used, the password becomes easy to crack, and the man-in-the-middle attack can be performed using the Wi-Fi link. Lack of secure encryption schemes throughout the chain of communication can cause such attacks. The authors of \cite{link2}, from IBM, have demonstrated the easiness of stealing a police quadcopter worth a thousand dollars by performing the man-in-the-middle attack. The researchers revealed that a hardware worth only $40$ dollars is sufficient to perform such an attack. This is a very clear example of an attack in which the controller is not even aware of the middle layer hacker.

\subsubsection{GPS Spoofing}
\begin{figure}[!t]
    \centering
    \includegraphics[width=90mm]{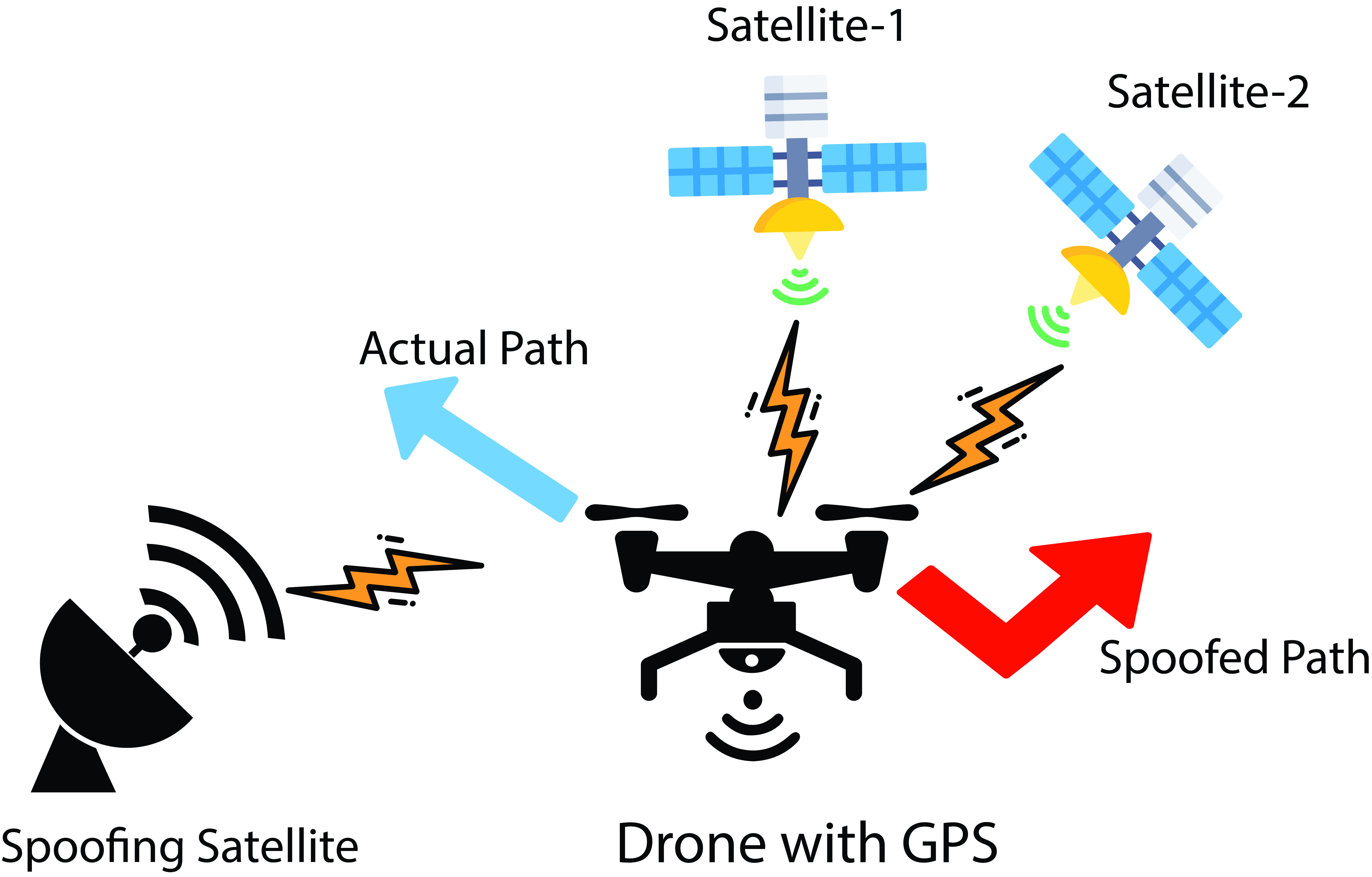}
    \caption{GPS spoofing attack on GPS Enabled Drone \cite{fig5}.}
    \label{spoofing}
\end{figure}

For communication, drones need incoming signals from GPS satellites, a two-way link between the drone and the ground-station, and signals notifying the drone's presence \cite{fig5}. Fig. \ref{spoofing} shows the basic working of the GPS spoofing attack in drone communication. Spoofing can be done using multiple transmitting antennas, in which the attackers transmitting antenna combines with the corresponding receiving antenna and transmits the false signals. In this process of getting the GPS coordinates of the drone, the drone is located by the satellite using GPS and its coordinates are then sent to the ground controller. The drones that do not have any encryption on their chipboard, are easily tracked by the hacker and they share a wrong location to the drone controller using a directional antenna with narrow beam-width aiming for the drone. GPS spoofing is mainly carried out on military drones as they are deployed at certain critical places that can provide highly confidential information about the other nations. It is relatively difficult to spoof military drones as they are highly equipped with encryption mechanisms. Spoofing can be done using multiple transmitting antennas \cite{spoof}, in which the attackers transmitting antenna combines with the corresponding receiving antenna and transmits the false signals. The spoofer can take the drone to any trajectory he/she wants without even giving the controller a hint as fake coordinates are sent to the controller at regular intervals. This technique can be used to reduce the velocity of the drone making it less useful. 

According to \cite{26}, on December $5$, $2011$, an American UAV was detected and shot down by Iranian forces near the city of Kashmar in northeastern Iran. According to the American officials, the UAV was spoofed and was forced to fly over Iran. The attackers hacked the UAV and injected it with wrong GPS coordinates. The incident resulted in a disturbance in the relationship between the two nations. The military drone was said to be using an inertial navigation system, and not the GPS navigation because of increasing number of spoofing and jamming attacks. Despite the measures taken to prevent any spoofing attack, or to protect the classified information available from the drones, the Iranians claimed that they could access it and reverse-engineered the entire drone.
 
The authors of \cite{spoof} have used Software Defined Radio (SDR) platforms to simulate GPS to transmit false signals to the target drone. This methodology has been used for a long time to hack or relay wrong information through drones. Using this approach, they divert and take control of the drones that depend on GPS for flight paths. For generating fake GPS signals, BladeRFx$40$ SDR is used, which is very versatile, and costs around $420$ dollars.

\subsubsection{Radar}
Mono-static radar is the most traditional way of searching for important entities. Similarly, it can be used to find drones. The radars send electromagnetic signals that can travel a long distance. These signals travel in all directions and wherever the presence of drone is detected, the signals reflect from the surface of the drone and are received at the other desired end. By further studying the signals, one can easily measure the velocity, direction, and altitude of the drone. A problem with this technique is that sometimes the electromagnetic signals consider obstacles like birds, airplanes, or kites as drones and transmit wrong information to the radar station, which in turn produces a loss. Radars operating in the millimeter wave (electromagnetic spectrum) range can be used for surveillance of small drones, even under adverse weather conditions, with high accuracy and with distance-independent resolution \cite{radar_em}. Moreover, to overcome these issues, hackers have tried to use various machine learning techniques, including the SVM classifier, binary decision tree, etc., to classify between the real drones and other objects \cite{radar}. In this technique, the detector is trained with a lot of data sets to distinguish between any obstacle and a drone. 

The authors of \cite{fraun} have discussed in detail about the ways in which the radars can be used to detect and identify drones. Various sensors such as optic or infrared sensors are also used to detect or identify drones. However, these sensors have various limitations in terms of range, and their reliability in night, rain, and fog. The use of radar is declared by the authors to be superior to visual optic sensors and infrared sensors because of their range. Applications of drones such as package delivery, and military operations, makes the drones very vulnerable to attacks that use radars. In all such areas, detecting and identifying the drones can be a threat to the drone itself, and might also result in the loss of other task-associated resources.

\subsubsection{Jammers}
\begin{figure}[!t]
    \centering
    \includegraphics[width=90mm]{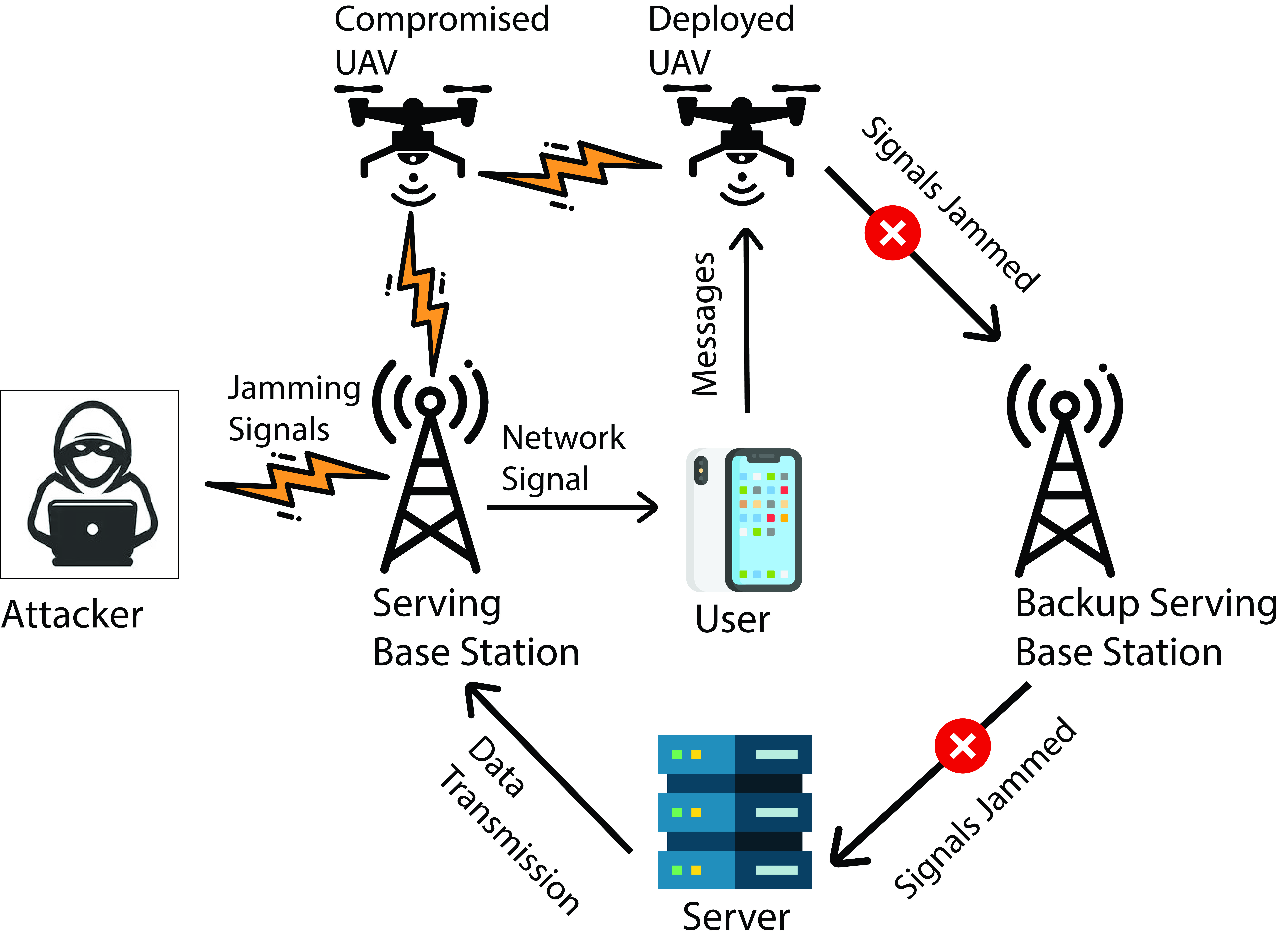}
    \caption{Jamming Attacks On Drone \cite{fig6}.}
    \label{jamming}
\end{figure}

These are the electronic devices used by the adversaries to block the signals at the receiver's end. It is mainly used for the disruption of communication between several users. It works on a simple principle in which a transmitter is used which is tuned to the same frequency as that of the target. If the jammer has enough power, it over-rides the frequency signals, thereby blocking every type of signal that the target can configure to. The jamming attack is analogous to the DoS attack, but the only difference is that in the DoS attack, the network, service, and the application layer get affected whereas in the jamming attack radio signals are used to attack the drones which mainly affect the physical layer. Signals of Wi-Fi and Bluetooth can be easily jammed, and that too using a low power jammer \cite{12}. The ability of the jammer is judged by its range. Jammers with a higher range can block signals of devices that are present upto that range. Fig. \ref{jamming} shows the implementation of the jamming attack. The attacker sends the jamming signal to the serving base station from his end, with the help of a UAV, which matches the frequency of the signal with the deployed drone. Thereafter, the signals between the drone and the backup serving base station are blocked. Hence, no data and commands are allowed to reach to the server, and the deployed drone becomes non-responsive thereafter.
Once the drone loses contact with the control station, many drones have a auto-pilot mode as a fail-safe which gets activated. Auto-pilot mode makes it easy for the attacker to launch a GPS-spoofing attack, and force a landing away from the original destination by spoofing the GPS signals \cite{jam_attack}. A technology was introduced in Australia a few years back, which would allow the hacker to commandeer a drone mid-flight and lands the drone in a defined exclusion zone by the new pilot \cite{jam_Aus}.

The authors of \cite{jammerinc} mentioned an incident in which GPS jamming was used to bring down $46$ drones during a show in Hong Kong. The drones started falling with great velocity. According to the board's executive director, these professional drones were equipped with fail-safe technologies to direct them back to their take-off location, but because the strength of the jamming signals was so strong, the drones started dropping mid-air only. The hacker had to just point the jamming device towards the drones and as soon as the signal interruption was detected by the drones, they started falling, as confirmed by Rex Ngan, founder of the Hong Kong Professional Unmanned Aerial Vehicles Association.

\subsubsection{Wormhole Attack}
UAV networks that utilize FANET or MANET are susceptible to routing-based attacks like wormhole. The communication between UAVs rely not only on the information exchange between UAV and the ground control station, but also amongst the UAVs. FANET uses a system of auto-configuration and self-healing to improve the reliability of the system, but it makes them vulnerable to wormhole attacks \cite{fanet_wormhole}. A wormhole attack is one of the most severe and grave attacks in MANETs.

In a wormhole attack, two attackers place themselves strategically in the network in order to listen-in on the communication between the drone-network. The attacker records the communication at one point in the drone-network and tunnels the information to the second attacker, where the recorded information is replayed. As the routing protocol algorithm looks for the nearest node to transfer the information, wormholes are placed so as to make the distant nodes believe that they are their closest neighbors. This kind of re-routing compromises the confidentiality of sensitive information, and also enables the attacker to launch an attack from any point in the network, because it practically controls all the routes discovered after the wormhole \cite{wormhole}.

Moreover, wormhole attacks in a UAV Ad Hoc Network (UAANET), made of a swarm of UAVs and a ground control station, are a high-level risk, and special attention needs to be paid to this multi-node attack. Even without the knowledge of any the cryptographic keys or hash-functions, the attacker is able to affect the integrity of the network by transferring control packets, and further, captures the data traffic \cite{uaanet_wormhole}.
 
\subsection{Potential Vulnerabilities In Different Drone Applications}
This section discusses the potential vulnerabilities in different drone application. Although the drone applications are vulnerable to every security issue mentioned above, we present the main security issue faced by the specific drone application. 

\subsubsection{\textit{Security Vulnerabilities in Mining Drones}}
Drones can help a lot in surface and underground mining \cite{mining}. Mining is a very tedious task and involves a risk to the life of the miners, so drones can be employed to decrease the workload and the risk to human life. Drones equipped with infrared night vision cameras can help in finding ores easily. They can also be equipped with a metal detector device to directly detect the ore even without a camera. Such applications may help in reducing the mining cost, and can increase the overall efficiency. There are various reasons and motivations due to which the adversaries may get attracted towards hacking such drones and launching a DoS attack. The other competitors may send unwanted requests to the mining drones, thereby preventing them from accurately identifying the ores and other valuables in the mines.

\subsubsection{\textit{Security Vulnerabilities in Disaster Management Drones}}
Drones can timely inform the respective emergency teams about the disaster situation and can help in preventing the loss of life and property. They can also help in delivering essential items to the disaster victims. Anticipated installation of drones can also be done in disaster-prone areas, to keep an eye on the upcoming disasters. The disasters may either be natural disasters like earthquake and flood, or man-made disasters like riots and terrorist attacks. Although there are multiple applications of drones in disaster anticipation, identification, prevention, recovery, and damage control, it is very important to safely deploy and use drones in such applications. These drones are highly vulnerable to de-authentication attacks. Both false positive and false negative messages from drones can result in major problems. Some hackers may try to de-authenticate the drones deployed for disaster anticipation and may send false positive messages to the emergency teams using their own malicious drones. This will end up in a waste of time and money for government bodies. False negatives may try to prevent the timely delivery of disaster-related messages to the emergency teams and thereby lead to loss of life and property.

\subsubsection{\textit{Security Vulnerabilities in Agriculture Drones}}
Various countries, like Sierra Leone, Somalia, and India, are dependent on agriculture for their living \cite{agri}. The farmers need techniques that help make farming easy and increase the productivity. Drones can be used for pollinating seeds which is a very important task for farmers to grow their crops. The drones can carry pollinating seeds and can have the dataset of the field where they have to sprinkle the seeds in the required quantity. This can decrease the work load of farmers and the mechanized sprinkling of seeds can also save seeds as drones can be programmed to sprinkle only the required quantity of seeds at appropriate locations. Drones can also help the farmers by spraying medicines in the correct quantity to kill unwanted plants like weeds and insects in the farms. In all the above discussed processes that are being carried out by the drone, the data should be accurate and if the drone gets hacked, the hacker can easily change the quantity of seeds or insecticides that have to be sprayed. The agricultural drones can be easily hijacked using the man-in-the-middle attack as the adversary can place himself between the drone and the ground controller and can manipulate the data already fed in the drone. Any unwanted change in the data can destroy the plants resulting in a great loss to the farmers, and as well as to the nation.

\subsubsection{\textit{Security Vulnerabilities in Military Drones}}
The initial drones were all very noisy, and therefore, it was difficult to use them for most of the hidden military operations. Various new drones have now been invented that make the least sound which makes them difficult to detect. The invention of such silent drones has increased their usability for various military purposes as they can fly up to a very remote location secretly. The cameras installed on the drones can be used to spot the enemy's location while carrying out any type of strike against the enemy. Although these drones have multiple advantages in military operations, if the drones are hacked or the communication link is spoofed, it can lead to disastrous results. The enemies can even hack and reprogram the drones to act against the army itself. The information leak by spoofing the communication link can also end up revealing military plans to the enemies. Therefore, it is very important to take care of all security standards before trying to deploy the drones for such critical applications. The most famous incident of such a case is also known as the Iran-U.S. RQ-170 incident \cite{26}, where Iran's military used GPS Spoofing to land a US UAV almost undamaged and reverse-engineered the entire design of the stealth drones to make their own Sa'egheh drones. This turned into an international incident.

\subsubsection{\textit{Security Vulnerabilities in Delivery Drones}}
Due to the increase in the pace of E-commerce, a lot of manpower is required especially for the last-mile delivery of products. The use of drones can be a promising solution as drones can deliver the products in less time and high accuracy. Drones can be used to deliver food, medicine, newspapers, and other things of daily basic needs. FAA approved the first NASA drone to deliver medicines in July $2015$. The UAV could successfully drop medicines to the health clinic in rural southwest Virginia \cite{7}. In $2016$, Amazon made its first drone delivery successful by delivering the package in $13$ minutes after being ordered by a customer in Cambridge \cite{8}. Amazon also launched a drone for delivery named Amazon Prime Air which can fly up to a range of $10$ miles for product delivery. These drones can take off and land autonomously, guided by GPS. Although, drones can help a lot in timely and accurate delivery, if these drones are hacked, it can end up in a big chaos. The hacker may use the radars to identify and capture the basic drones used for delivery and may guide the drones to deliver packets to different destinations or to himself. Therefore, care of the security standards is important even for the delivery drones. Since a huge number of people are using e-commerce, any misstep could endanger the privacy of billions of people in the future.

\subsubsection{\textit{Security Vulnerabilities in Drones for Urban Planning}}
Urbanization refers to the heavy movement of people from rural areas, like villages, to the cities. Drones can be highly helpful for architects to take some major decisions regarding renovations and new constructions. Drones can also help in making and analyzing the plans for water management in cities. A drone can be deployed with GIS (Geographic Information System), by which drones can easily capture, analyze and manipulate geographical data of the water supply management. It is important to have well-defined security measures for such drones as well. Various rules have to be followed to approve a safe construction. There have been various cases of buildings getting collapsed due to illegal construction resulting in a loss of life. If the architects rely on the results submitted by drones, and the values calculated and submitted by drones are not secure, then illegitimate people might try to hack and manipulate the drone functioning to get their illegal constructions approved. To conceal their identity, people with ill intent could crash such drones which could lead to a loss of resources. Other owners of illegal or unauthorized construction sites may deploy jammers to prevent such drones from identifying illegal constructions.

\subsection{\textcolor{black}{Classification of Drone Communication Systems}}
\subsubsection{\textcolor{black}{Drone-to-Drone}}
\textcolor{black}{Even though drone-to-drone (D2D) communication has not been standardized yet, it can be seen as a peer-to-peer (P2P) network \cite{d2d_security}. This makes D2D communications susceptible to various P2P attacks (DoS/DDoS, jamming attack, etc.).}

\subsubsection{\textcolor{black}{Drone-to-Infrastructure}}
\textcolor{black}{Drone-to-infrastructure communication can be further classified into categories such as:
\begin{enumerate}
    \item Drone-to-Satellite: This infrastructure is used for the drone to coordinate with the GPS. Although, it is expensive to set-up and maintain, such communication systems are considered safe and secure.
    \item Drone-to-Network: This type of communication is useful for cellular networks (4G, 5G, etc.), and it is very important to ensure their security when used.
    \item Drone-to-Ground Station: This infrastructure is based on common wireless technologies like Bluetooth and Wi-Fi. They are public, and hence not secure, making them very susceptible to man-in-the-middle attacks and eavesdropping.
\end{enumerate}}

\par
 \textbf{       Summary:}
This section discusses the security issues being faced by the existing drone applications. Many attacks like DoS attacks, De-authentication attacks, Man-In-The-Middle attacks, that deal with the tampering of the data in the drones are mentioned in this section. Several other attacks that deal with the position of the drones like GPS spoofing, jamming attacks, radars are also discussed. In the next section, we discuss the overview and fundamentals of various emerging technologies such as blockchain, SDN, ML, and fog computing that can help in preventing the above-mentioned attacks on the drone applications. \textcolor{black}{Furthermore, we gave a brief classification distinction between drone communication systems.}

\section{Overview and Fundamentals of Various
Emerging technologies for Secure Drone Communication}
\label{sec3}In this section, we discuss the four main emerging technologies that are being widely used and explored for making drone communication fast, reliable, and secure. Mainly we discuss the use of blockchain, ML, SDN, and fog computing for secure drone communication. 
 \begin{figure*}[!t]
 	\centering

 	\includegraphics[width=180mm]{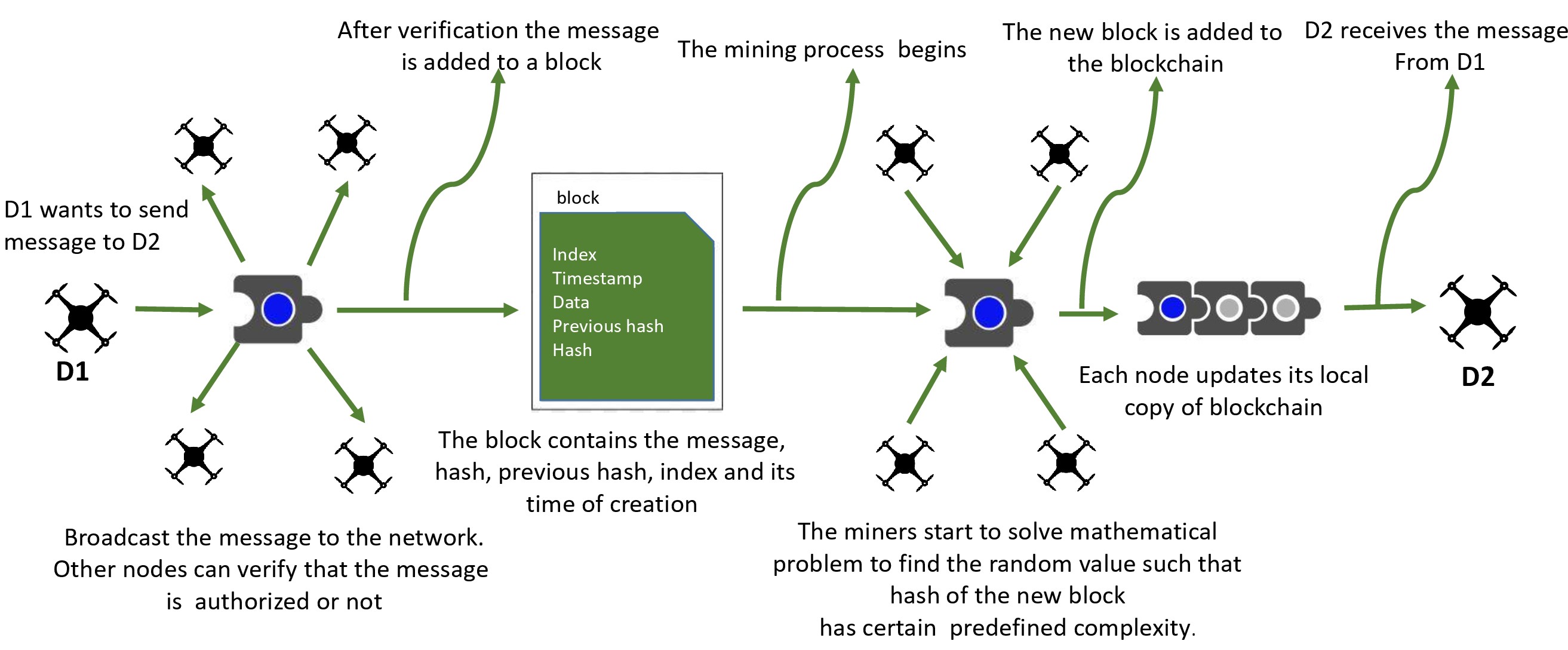}
 	\caption{Working process of blockchain.}
 	\label{bcworking}
 \end{figure*}
\subsection{Drone Communication Architecture using Blockchain}
According to FAA, $1.3$ million drones have been registered with the FAA in $2019$ and the number is expected to increase to $7$ million by the end of $2020$ \cite{dronenumber}. With the rapid increase in the number of drones, the data generated by them is also increasing rapidly, which has increased security concerns about the data. Researchers state that blockchain can contribute to another layer of security to drone communication, which would prevent data retrieval and tamper by unauthorized persons \cite{cov}. Furthermore, the data on the blockchain is distributed, such that it becomes very difficult for an adversary to hack a single system to get control over the complete data in the network. Figure \ref{bcworking} shows the basic working process of the blockchain technology.
 
\subsubsection*{\textit{Motivation For Using Blockchain For Drone Communication Security}}
A blockchain is a growing chain of blocks linked to each other using cryptographic hash functions. The drone applications are becoming highly popular and are gradually being used in almost all domains and spheres of life. With the increasing number of drone applications, it is imperative to keep the transactions between the drones and other users secure, cost-effective, and privacy-preserving. The blockchain technology is a highly promising solution that can be used to deploy real-time drone applications. Once a transaction is recorded on the blockchain, it remains immutable and no adversary can try to tamper the records \cite{blocksolauth}. Furthermore, the use of smart contracts can help a lot in performing different transactions between different parties in a secure and cost-effective manner. Depending on the nature of the application, different kinds of blockchain networks can be created, such as public, private, consortium, and hybrid. Moreover, there is a vast variety of consensus algorithms used in the blockchain network ranging from PoW, PoS, PoB, DAG, and so on. All these features of blockchain can be leveraged to make drone communication secure, reliable, and cost-effective. In Section \ref{sec5}, we discuss in detail the various blockchain-based models to secure drone communication.
  \begin{figure}[!b]
    \centering
    \includegraphics[width=90mm]{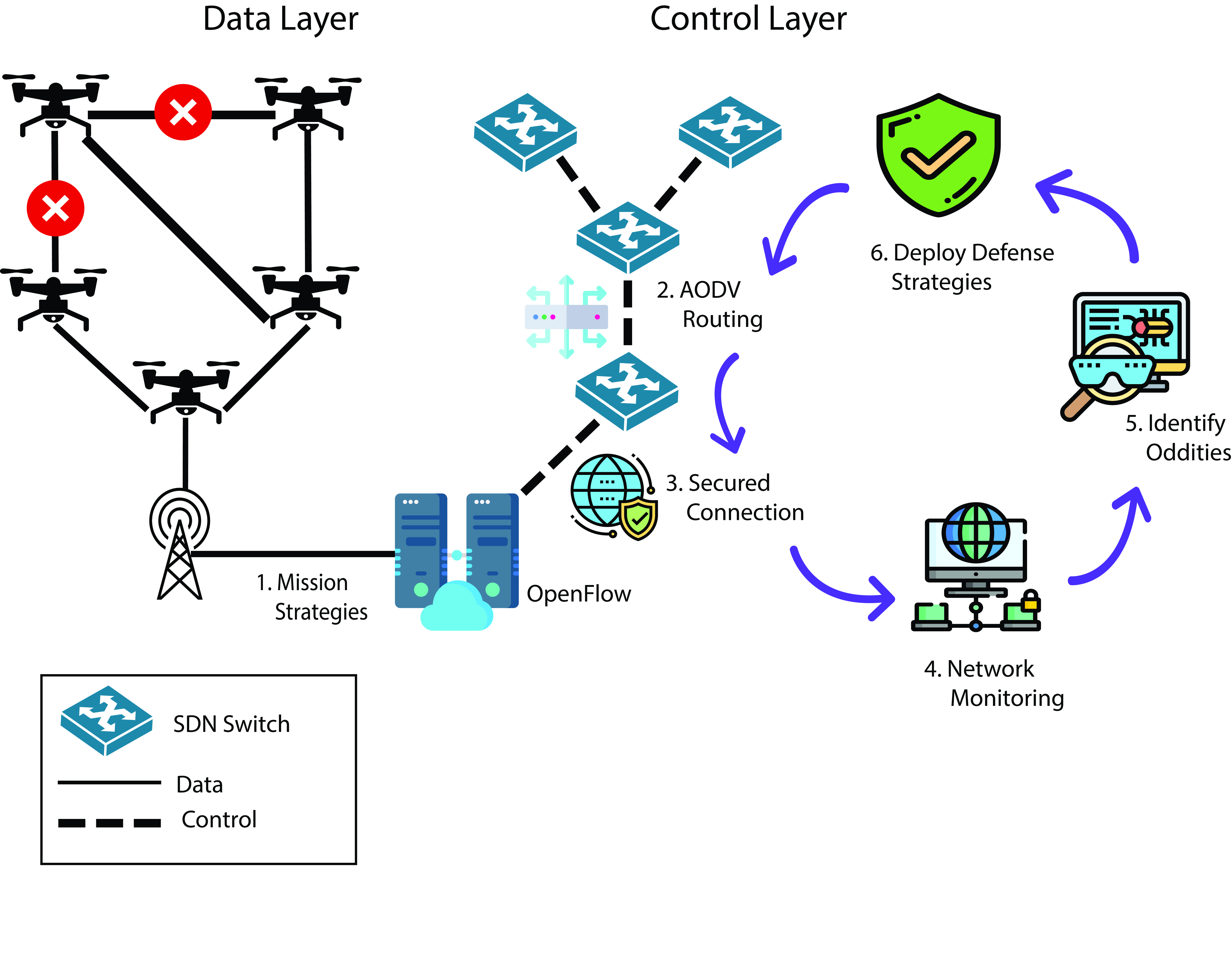}
    \caption{Basic SDN Architecture \cite{fig11}.}
    \label{sdn_arch}
\end{figure}
\subsection{Drone Communication Architecture Using SDN}
Software-Defined Networking (SDN) is a networking architecture that can be used to control or program the network centrally using software applications. SDN helps in the consistent management of the network as everything in the network is centrally programmed. The architecture of a typical SDN-based drone communication network is shown in Fig. \ref{sdn_arch}. The figure illustrates a simple use case of SDN in drone applications. It shows the steps involved in transmitting data from the drones in the data plane/layer to the control plane for getting the data processed and then getting back the required output. In a typical SDN-based drone communication network, each drone in the network behaves as an individual switch. The application plane of the SDN sits on a centralized controller and is responsible for the implementation of any and all high-level functions to be performed by the network as a whole. The centralized controller also houses the control plane, which would command and control data flows between the drones. The data plane consists of the drones themselves, which respond to commands from the controller. There exist a variety of protocols and standards for performing various functions in the network. Because the SDN architecture decouples the control plane from the data plane, protocols in different planes can be implemented independently. This offers a considerable degree of freedom in the design of an SDN-based drone communication network. The authors of \cite{rev3} review the various 5G techniques based on UAV platforms using network layer techniques like software-defined UAV networks.

\subsubsection*{\textit{Motivation For Using SDN For Drone Communication Security}}
As discussed above, SDN enables the network to be centrally controlled, which makes the network reliable. Moreover, SDN's decoupled data layer, control layer, and application layer makes the network control directly-programmable. The increasing drone applications make use of real-time video streaming which can be achieved by the use of SDN, as it can provide better QoS because the traffic is automatically controlled in the SDN. The drone is highly resource-constrained so there are many vulnerabilities that can be prevented by the use of SDN, as the controller can keep a close check on the data traffic. The above-mentioned parameters of SDN helps in maintaining the overall security in the network. Section \ref{sec6} discusses the SDN-based DCN models that can help in making the drone communication secure from different types of attacks.

\subsection{Drone Communication Architecture Using Machine-Learning}
\begin{figure}[!t]
 	\centering

 	\includegraphics[width=90mm]{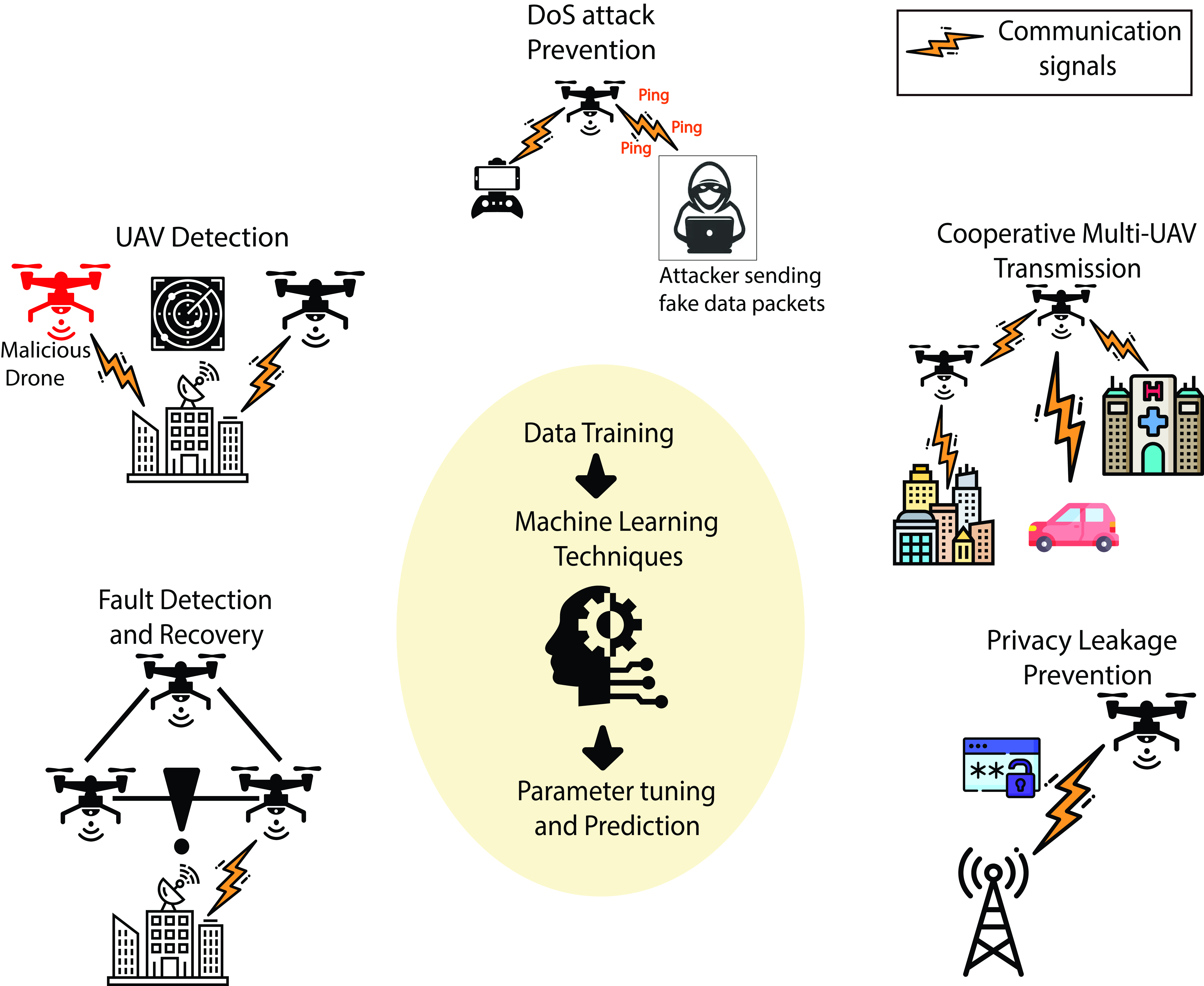}
 	\caption{Basic ML Techniques Architecture \cite{mlfigure}.}
 	\label{mlfigure}
 \end{figure}
Machine learning is a technique that provides the system with the ability to learn automatically and ameliorate using past experiences without being explicitly programmed. Once the data is fed, ML learns and predicts the output automatically without much human intervention. The ML algorithms need a large amount of training data to make more accurate predictions. ML algorithms are broadly divided into two categories, i.e., supervised machine learning (training dataset can be classified into several labels) and unsupervised machine learning (training data is not classified) algorithms. Figure \ref{mlfigure} shows the basic architecture of ML-based drone communication applications. The figure illustrates the various ways in which the ML techniques can assist in making the drone communication secure. Several ML algorithms like CNN (Convolutional Neural Network), SVM (Support-Vector Machine), ANN (Artificial Neural Network), RNN (Recurrent Neural Network), etc. can be used for making drone communication secure. ML algorithms, such as LSTM (Long Short-Term Memory), can also be used for detecting the faults in drone communication, and the recovery methods are sent to the drone for its safety. A classification algorithm can be applied which can help in detecting the DoS attacks and other attacks that make use of the fake and affected data packets to paralyze the network. The data packets can be easily classified as either benign or affected packets, which can prevent the network from getting hacked. These diverse applications of ML can help achieve highly secure drone communication.

\subsubsection*{\textit{Motivation For Using ML For Drone Communication Security}}
The ML algorithms learn from the training data and improve themselves for achieving better results and high accuracy without any human intervention, which is a huge advantage. The ML algorithms can be deployed for detecting the presence of malicious drones in the network and can help in preventing the attacks such as man-in-the-middle attack \cite{mitmml} and spoofing attacks \cite{mlspoof}. Such algorithms keep on improving with increasing experience, and provide better and more accurate results. The models can also be trained to automatically detect and recover from the faults using neural networks and LSTM. Moreover, ML algorithms can handle multi-dimensional and diverse data. All these parameters make these algorithms highly suitable for use in drone applications. In Section \ref{sec7} we discuss in detail the use of various ML techniques to secure drone communication. 

\subsection{Drone Communication Architecture Using Fog Computing}
\begin{figure}[!t]
    \centering
    \includegraphics[scale=0.9]{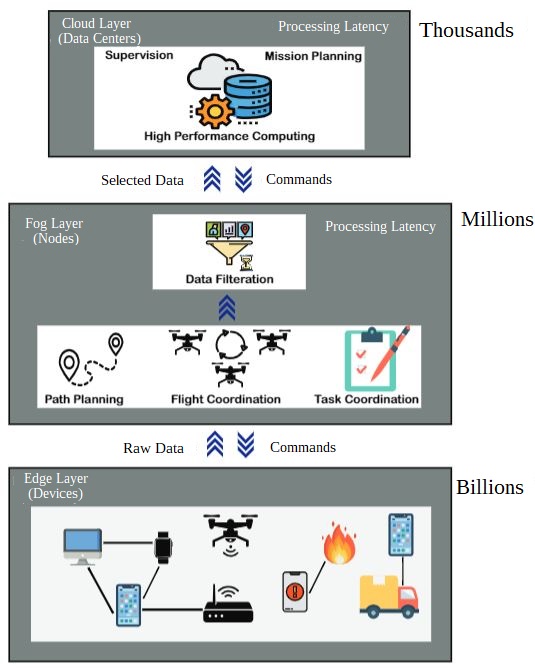}
    \caption{\textcolor{black}{Layered Architecture of Cloud, Edge, and Fog Computing \cite{fig12}.}}
    \label{foglayer}
\end{figure}

The concept of fog computing was first introduced by CISCO in $2014$ \cite{ff}. Fog is considered to be a dimension that extends the use and capabilities of the cloud. Fog is not particularly a substitute for cloud computing; instead, it is a large complement of cloud computing. The fog layer is a layer or stratum between the edge devices and the cloud. Deploying cloud servers is very difficult as it is very costly and is very difficult to establish. So a new concept came into the market in $2014$, by which the load of the cloud can be minimized. Fog is a smaller version of the cloud which can be placed near to the end devices. Fig. \ref{foglayer} shows the \textcolor{black}{layered architecture of cloud, edge, and fog computing combined}. What happens in fog computing is that whenever an end device user requests some query of fetching any data or uploading any data, the mobile network helps in connecting to the nearest fog node available. Now the data can be easily fetched and stored in the fog. The fog uses LAN (local area network) whereas for accessing cloud facilities we need to access the internet through WAN (wide access network) which will take more time as well as more cost. So, fog computing is very helpful in many aspects like cost, time, and security. Fog domain is made by combining multiple fog nodes which can be switches, gateways, routers, smartphones, or PCs to form the fog stratum.

\subsubsection*{\textit{Motivation For Using Fog Computing For Drone Communication Security}}
Fog Computing is a paradigm that can help in processing and accessing large data rapidly, efficiently, and with the least possible latency. This is a layer between the end-device and the cloud servers. Fog computing helps in increasing the QoS and QoE, as the retrieval time of the data is very less in the fog. Fog computing also reduces the data load on the cloud servers and makes the data dissemination very cost-effective as well as reliable. The fog is a decentralized paradigm in which the data is stored at multiple fog layers. This proves advantageous as compared to when data is stored in one place, because data in the fog has no central entity handling the entire data making it less vulnerable. This also prevents the cloud server from being getting affected, as the vulnerability can be detected at earlier stages. These aspects make fog computing a very important technology in making drone communication secure. The fog-based DCN methods and models that help make the drone communication secure are discussed in Section \ref{sec8}.

 \begin{figure*}[!t]
 	\centering
 	\includegraphics[width=180mm]{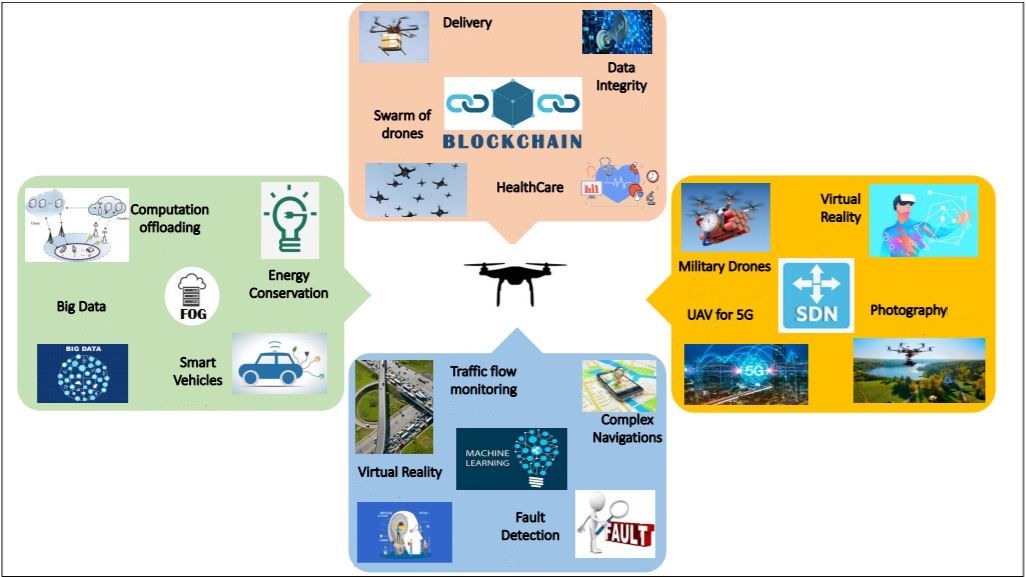}
 	\caption{Different drone applications using various security techniques.}
 	\label{drones}
 \end{figure*}
 
Fig. \ref{drones} shows various drone applications in different domains that have used \textcolor{black}{blockchain \cite{BDI4}, SDN \cite{intent}, ML \cite{ML1}, or fog computing \cite{fogres} for securing drone communication.} The rest of this paper discusses the usage, and benefits of these technologies in making drone communication more secure in detail.

\section{Applications of blockchain for Drone Communication Security}
\label{sec5} Drone technology has been there for almost a century, but in the recent past, it has gained importance in fields such as agriculture, security, wildlife conservation, delivery, and so on. Blockchain technology is said to have the potential to improve data security and transparency across multiple domains \cite{newly3,newly4,newly5,newly7}. In this section, we will elaborate the models and the mechanisms based on blockchain that can be used in making the drone communication secure.

Various other non-blockchain technologies have also been proposed to increase drone security. There are various issues related to such solutions that can be resolved using the features of blockchain technology. A model proposed in \cite{non1} helps in maintaining data integrity by using sensor Physical Unclonable Function (PUF). This method provides data integrity but fails in maintaining self-trust and the data provenance. Another non-blockchain model for preventing a wormhole attack has been discussed in \cite{worm}. The authors use a label-based method for detecting the attack. The model only addresses wormhole attacks and is still vulnerable to other attacks that blockchain could prevent such as DDoS attacks, and GPS spoofing attacks.

\begin{figure*}[!t]
    \centering
    \includegraphics[width=180mm]{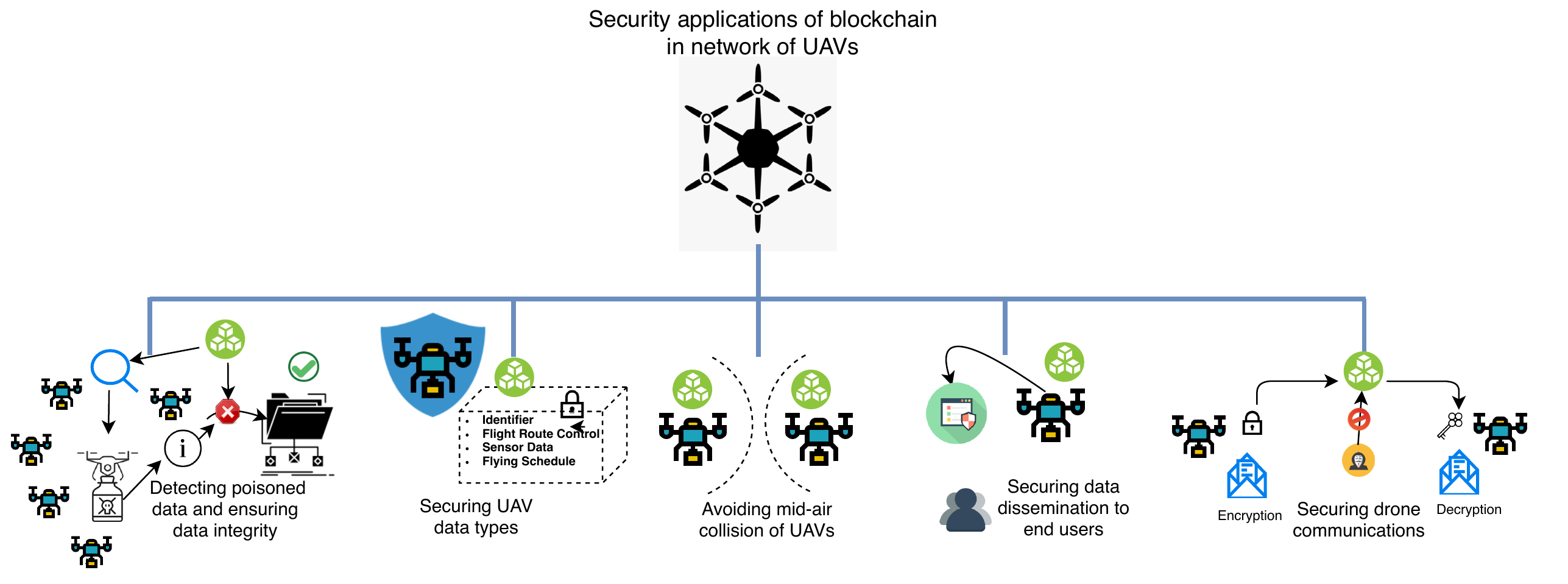}
    \caption{\textcolor{black}{Security applications of blockchain in a network of UAVs.}}
    \label{blockchainindroneapplications}
\end{figure*}

We further discuss the list of specific security issues that can be resolved and prevented using blockchain as a solution. Fig. \ref{blockchainindroneapplications} shows the various security applications of blockchain in a network of drones.
  

\subsection{ \textit{\textbf{\textcolor{black}{Air Traffic Management}}}}
UAVs have recently gained huge popularity. Thus, with a number of drones, their paths may cross with one another and sudden collision may occur. Therefore, it is necessary to devise a solution and a platform to maintain optimal paths for air traffic management \cite{newly10}. Such traffic management in drones is different from traditional road traffic management, as there is no well-defined path of travel for each drone and the coordinates need to be maintained in $3$ dimensions. Blockchain and IoT have many advantages over traditional internet-based systems due to the fact that the internet-based systems are more prone to cyber-attacks that would degrade or disrupt the functioning of the drones. Traditionally, GPS coordinates are used for UAV localization and avoiding traffic violations. However, such approaches are difficult to apply for complex paths due to pilot errors and other intrusion attacks. The authors of \cite{NBTM} suggest that the neural-blockchain based transport model (NBTM) can significantly help in optimizing the problem of the air traffic violation. This model involves the use of $3$ different blockchain networks to form a master blockchain, taking the input parameters as a function of the reliability of connections and reliability of flyby time. The model also generates feedback for initial inputs while iterating towards an optimal solution. The forward propagation is done between Blockchain $A$ and $B$, and backward propagation is between Blockchain $C$ and $D$. The primary components of NBTM are blockchain and neural networks. The neural model is a 4-layer network having $B$ and $C$ as intermediate layers. The output of the neural network model is used to form the optimal path for UAV to travel. This model does not employ any separate mechanism for security, but is simply dependent on the basic principles of the blockchain. The simulation results demonstrate that the proposed neural-blockchain enhances the reliability (the statistical parameter for evaluation of consistency) of the network with a lesser failure rate. Due to the availability of a feedback mechanism, the model reduces the computation power demand, resulting in lesser complexity and yields higher efficiency when compared with the model proposed in \cite{NBTM2}. In \cite{NBTM2}, the authors have proposed a model for reducing the number of transaction required for updating the ledger in the Internet of Vehicles (IoV), so that less latency is experienced in maintaining the air traffic whereas in \cite{NBTM}, feedback mechanism gives better results. However, the dynamic partitioning between a centralized system and the blockchain-based system is still a challenge to be worked upon.

\textit{Preventing mid-air collisions: }
Air traffic Control (ATC) needs improvement in preventing mid-air collisions due to the increasing number of UAVs. The Las Vegas incident wouldn't have happened if proper precautions were taken \cite{2}. Due to resource constraints and heavy traffic, the UAVs are subjected to transmission delays in communication with the ground station, unlike high-speed LAN serial communication, and therefore an innovative method to improve ATC and prevent mid-air collisions is required. The authors of \cite{BDI3} propose a blockchain-based solution for ATC management to prevent mid-air collisions. Similar to \cite{NBTM} and \cite{BD2}, the authors of \cite{BDI3} focus on physical protection of drones in case of high air traffic. However, different from \cite{NBTM} and \cite{BD2}, the authors of \cite{BDI3} focus more on exploiting the feature of peer-to-peer transactions in blockchain technology as compared to the feature of tamper-less data storage. If the path at which the drone has to traverse is defined and is stored in a tamper-less data storage, then the drone can be made secure, as no adversary can change the path of the drone for its benefit. Because of drones' agility, collision avoidance algorithms can also be employed to prevent mid-air collisions. A fast obstacle collision avoidance algorithm for UAVs has been proposed by the authors of \cite{midair_algo}. Using this algorithm, the drone can avoid static and dynamic obstacles, while still being able to get back on it's initial trajectory. Mid-air collisions can be dangerous as after the collision the drone could fall down to the ground and can cause harm to human life. Similarly, mid-air collisions could also bring in various threats to the airplane flying in the sky as any collision with the airplane would lead to loss of lives as well. In the proposed model, blockchain is used to store UAVnet data that comprises UAV-ID, flight route sheet, flying schedule, and sensor’s data. The computing UAVs are divided into $m$ groups each containing $n$ number of UAVs. Out of those $m$ groups, one is used to store information broadcasted from other UAVs and acts as actual blockchain participant. The other computing UAVs simulate the possible paths for the idle UAV to reach its destination. The optimal path that would limit the mid-air collisions is chosen by the Proof of Graph (PoG) consensus mechanism which is based on Simplified Memory Bounded A* (SMA*) Algorithm \cite{bound}. The authors compare the SMA* Algorithm with A* and Dijkstra’s algorithms. Although Dijkstra can find the shortest path, the algorithm must explore all possible paths, resulting in high complexity. A* algorithm uses exponential memory whereas SMA* uses bounded memory, where in exponential memory, addition of data increases the computation time exponentially \cite{exp}, and in bounded memory, the computation time depends on the amount of memory the data needs \cite{bounded}. This specification of the bounded memory makes the algorithm memory efficient and reduces the required computation time.

\subsection{ \textit{\textbf{\textcolor{black}{Geo-fencing System}}}}
Geofencing can be defined as the virtual fencing or boundary created to disengage UAVs from entering a sensitive area such as prisons, airports, and private properties \cite{geo}. It is similar to road networks where some vehicles are not allowed to enter certain zones. However, creating fences is more complicated in IoD due to lack of well-defined pathway, and motion being in $3$ dimensions. Traditionally, DJI’s GEO System was used to mark where it is safe for the drone to fly and where the authorities may raise concerns about the flight of the drone\cite{DJI}. However, such systems may not be suitable for drones, as there are certain prohibited zones where the drones cannot fly, like near the airports \cite{airport}. These systems give the optimal path which cannot be used in the case of the drones as the proposed optimal path may lie in the prohibited zone. Blockchain can be effective in maintaining such a restriction based on the $3$ dimensional coordinate system in real-time. The pioneer work for blockchain-based flight space allocation in a decentralized fashion is \cite{BD2}. Unlike \cite{NBTM}, the authors of \cite{BD2} focus mainly on preventing the entry of drones into restricted areas rather than performing complete traffic management. In the proposed model, the UAV during its flight adds its request for air-space to the blockchain network. The trajectories are then added in such a way that it does not conflict with any restricted zone defined through virtual fencing. Blockchain can maintain the constraint that the optimal path should be chosen such that it should not lie in the prohibited flying zone. This also mandates that air paths of multiple UAVs do not cross with one another which would eventually lead into a crash. The proposed model is better than the baseline scheme in \cite{NBTM}, as the proposed model uses blockchain both for geo-fencing and avoiding traffic violations. The benefits of using blockchain to maintain geo-fencing include its immutability and safety from cyber attacks. However, as transactions continue and records grow, and block sizes increase in a blockchain, eventually exceeding any limits set, each transaction will need more time to be processed. Thus there is a need for blockchains with a higher TPS (Transactions per Second) rate to avoid bulking. Some newer blockchain-based structures that offer higher transaction rates of $3500$ TPS have also been proposed in the literature \cite{hedera}.

\subsection{ \textit{\textbf{\textcolor{black}{Maintaining data integrity}}}}
The immense data including the geographical addresses, and the data captured by the sensors on the drones, can be collectively used to profile an individual, and thus, can lead to privacy leakage \cite{IoD}. The Iranian government claimed to be able to access all the information from the American UAV and reverse-engineer the entire drone \cite{26}. As drones have limited computation resources, the data processing can be done in the cloud. In traditional cloud-based solutions, the Zone Service Providers (ZSPs) provide for the navigation of the drones and the feedback systems between drones. However, ZSPs are vulnerable to attacks due to high latency and high false rate \cite{IoD}. An efficient IoD using blockchain technology is proposed in \cite{BDI4}. Different from \cite{NBTM} and \cite{BD2}, the focus of the authors of \cite{BDI4} is more towards securing the important data being sent by drones, rather than physically preventing them from colliding or entering restricted areas. Tamper-proof and transparent storage of data are the main features of blockchain technology being exploited by the authors of \cite{BDI4}. In the proposed algorithm, firstly, the drone enrolls itself in the blockchain ledger for storing the data, and a unique ID is assigned to the drones. The data is then hashed for maintaining the integrity and is then uploaded to the blockchain network, via the controller. After the data is successfully uploaded, an acknowledgment is sent to the drone. The data records are transformed into a Merkle Hash tree \cite{merkle}. Furthermore, data auditing is done in the cloud which is a crucial step as it helps to detect any anomaly in the data. The proposed model was analyzed for the response time with varying numbers of drones. The simulation results demonstrate that the response time increases linearly from about $400$ms for $100$ drones to about $550$ms for $1000$ drones, thus providing better scalability. The average response time latency is also fairly stable, varying from $350$ms to $1000$ms for a $100$ drone network. The proposed model also makes the network less vulnerable to attacks like DDoS attacks and data losses, while making it more accountable. One major challenge is the time delay in the drones due to proof of work. For mining a block, computing machines require enough time, which results in latency \cite{mine}. Due to hardware constraints in drones, lightweight cryptography and DAG-chain based consensus algorithms can be developed \cite{my3}.

\textit{\textbf{Secure Data Dissemination: }}
Data dissemination is the process of distribution of data/information to its end-users. The authors of \cite{BDI2} propose a blockchain-based algorithm which helps in secure data dissemination in the IoD (Internet of Drones). Although the model presented in \cite{BDI2} is based on blockchain, it is not designed to keep the localization information as shown in \cite{local}. The authors of \cite{BDI2} make use of blockchain technology only to secure the data transfer between the drones and drone controllers. A combination of the approaches discussed in \cite{local} and \cite{BDI2} would be a promising solution for securing both localization and data dissemination. The proposed model in \cite{BDI2} is designed using three layers, namely, the user layer, the infrastructure layer, and the IoD layer. In the user layer, blockchain technology is used for the verification and the security of each transaction made in the model. The second layer, the infrastructure layer, consists of all the base stations, which ensure the connectivity between the drone controller, the drone, and the end-users. IoD layer consists of the drones that capture real-time data and communicate amongst themselves for making certain decisions. Two types of nodes are considered in this model, (i) forger nodes and (ii) normal nodes. The forger node is used for creating new blocks in the blockchain, whereas, the normal nodes are used for the verification of the blocks in the blockchain. This model works in three stages. First, the forger node is selected, and the other remaining nodes are declared as normal nodes. After the forger node is selected, the hash value is calculated by the forger node using the PoS (Proof of Stake) consensus algorithm \cite{PoS}. The other nodes validate the hash value broadcasted by the forger node by comparing it with the hash value that is generated using the Merkle Hash tree. If both the hash keys match, the block is validated and is added to the main chain. The forger node then encrypts the data packets and sends the request to the public distributed blockchain. When the request is accepted, the forger node computes the digital signature of the data packets with its private key and broadcasts it to the public blockchain. The data is stored in the blockchain and can be accessed only using the decryption key, so attacks like spoofing or DoS attacks can be prevented using this algorithm. The authors evaluate the security of the proposed model in terms of communication cost and time. The simulation results demonstrate that the proposed model provides data authentication, authorization, and accountability that is not offered by other state-of-the-art related works. Another work related to securing data dissemination in IoD is \cite{IoD}. The authors of \cite{IoD} use Identity-based encryption techniques for secure data dissemination. Such techniques can provide data integrity and identity anonymity only, and fail to provide authentication, authorization, and accountability of nodes in the network. Also, there is no proposal for data verification and validation in \cite{IoD} as compared to the blockchain-based approach used in \cite{BDI2}.

\begin{table*}[]
\centering
\caption{A summary of advantages and disadvantages of major applications of blockchain for drone communication security.}
\begin{tabular}{|l|l|l|l|}
\hline \rowcolor[gray]{0.8}
\begin{tabular}[c]{@{}l@{}}Major\\ approaches\end{tabular} & Advantages & Disadvantages & \begin{tabular}[c]{@{}l@{}} \textcolor{black}{Benefits over}\\ \textcolor{black}{traditional approaches} \end{tabular}\\
\hline
\cite{NBTM} & \begin{tabular}[c]{@{}l@{}}• Significantly helps in optimizing the \\problem of air traffic violation\end{tabular}               & \begin{tabular}[c]{@{}l@{}}• Does not support dynamic\\ partitioning of UAVs into groups\end{tabular} & \textcolor{black}{Reduced Latency}\\
\hline
\cite{BD2}                                & \begin{tabular}[c]{@{}l@{}}• Prevents the UAVs from entering into\\ any restricted zone through virtual   fencing\end{tabular}               & \begin{tabular}[c]{@{}l@{}}• Can support only a limited\\ number of transactions per minute\end{tabular} & \begin{tabular}[c]{@{}l@{}} \textcolor{black}{Immutability, Safety}\\ \textcolor{black}{from cyber attacks} \end{tabular}\\ \hline
\cite{BDI4}                               & \begin{tabular}[c]{@{}l@{}}• Supports high scalability with stable\\ response time latency\end{tabular}                                       & \begin{tabular}[c]{@{}l@{}}• Significant latency in the data \\ transmission\end{tabular} & \begin{tabular}[c]{@{}l@{}}\textcolor{black}{Scalabilty, Data }\\ \textcolor{black}{integrity} \end{tabular}\\ \hline
\cite{BDI3}                               & \begin{tabular}[c]{@{}l@{}}• Supports high computation speed and\\ memory efficiency with SMA*\end{tabular}                                   & \begin{tabular}[c]{@{}l@{}}• Does   not support dynamic\\ partitioning of UAVs into groups \end{tabular} & \begin{tabular}[c]{@{}l@{}}\textcolor{black}{Cost-effective,}\\ \textcolor{black}{Scalability} \end{tabular}\\ \hline
\cite{local}                              & \begin{tabular}[c]{@{}l@{}}• The chances of localization errors are\\ reduced to $1/4^{th}$ \end{tabular}                                   & • Suspectible to $51$\% attack & \textcolor{black}{Localization}   \\ 
\hline
\cite{BDI2}                             & \begin{tabular}[c]{@{}l@{}}• Proof-of-stake consensus algorithm is\\ used to significantly reduce the\\ computation time and cost\end{tabular} & \begin{tabular}[c]{@{}l@{}}• Regulatory control and\\ governance features are missing\end{tabular} & \begin{tabular}[c]{@{}l@{}}\textcolor{black}{Authentication, Authoriz-}\\ \textcolor{black}{ation, Accountability} \end{tabular}     \\ \hline
\end{tabular}
\label{bloadv}
\end{table*}

\begin{table*}[]
\centering
\caption{Applications of Blockchain for drone communication security. }
\begin{tabular}{|p{0.033\linewidth}|p{0.11\linewidth}|p{0.24\linewidth}|p{0.11\linewidth}|p{0.21\linewidth}|p{0.15\linewidth}|}
\hline \rowcolor[gray]{0.7}
Ref.                       & Attack                                                                                              & Mechanism                                                                                                                                                                                            & \begin{tabular}[c]{@{}l@{}}Blockchain\\ Feature used\end{tabular}                                              & Major achievement                                                                                                                                                                                & Open issues                                                                                                                              \\ \hline
\cite{NBTM}    & \begin{tabular}[c]{@{}l@{}}GPS Spoofing,\\Jamming\\attack\end{tabular}                             & \begin{tabular}[c]{@{}l@{}}The optimal path generated\\by the neural network is\\stored in the blockchain\end{tabular}                                                                            & \begin{tabular}[c]{@{}l@{}}Peer-to-Peer\\model \end{tabular}                               & \begin{tabular}[c]{@{}l@{}}The model gives the best\\ path with a maximum\\ failure rate of $25.8$\%\end{tabular}                                                                                & \begin{tabular}[c]{@{}l@{}}Making the model\\ efficient for a large\\ number of UAVs\end{tabular}                                        \\ \hline \rowcolor[gray]{0.9}
\cite{BD2}     & \begin{tabular}[c]{@{}l@{}}GPS spoofing, \\ DoS attacks,\\ DDoS attacks\end{tabular}                & \begin{tabular}[c]{@{}l@{}}The blockchain assigns \\ trajectory to the UAVs such \\ that no route clashes with\\ the other UAVs route.\end{tabular}                                                  & \begin{tabular}[c]{@{}l@{}}Tamper-proof\\data,\\ Peer-to-Peer \\ network\end{tabular}                            & \begin{tabular}[c]{@{}l@{}}A collision-free trajectory\\ is proposed such that the\\ UAV does not enter the\\ geo-fenced zone\end{tabular}                                                       & \begin{tabular}[c]{@{}l@{}}Model has to be\\ trained for handling\\ large number of\\ UAVs\end{tabular}                                  \\ \hline
\cite{BDI4}    & \begin{tabular}[c]{@{}l@{}}DoS attacks,\\ DDoS attacks\end{tabular}                                 & \begin{tabular}[c]{@{}l@{}}The data generated from \\ the sensors is stored in the \\ Merkle Hash tree which\\ ensures data integrity\end{tabular}                                                   & \begin{tabular}[c]{@{}l@{}}Distributed-\\ database,\\ Public key\\ infrastructure\end{tabular}                  & \begin{tabular}[c]{@{}l@{}}The average response\\ time for data transmission\\ with $1000$ drones is\\ $550$ms\end{tabular}                                                                      & \begin{tabular}[c]{@{}l@{}}Implementing\\ private blockchain\\ can make the\\ system more secure\end{tabular}                            \\ \hline \rowcolor[gray]{0.9}
\cite{BDI3}    & \begin{tabular}[c]{@{}l@{}}Man-in-the-\\ middle attacks,\\ GPS spoofing,\\ DoS attacks\end{tabular} & \begin{tabular}[c]{@{}l@{}}The details of the UAV stored \\ in the blockchain are used to \\ calculate the optimal path\\  using the SMA* algorithm\end{tabular}                                     & \begin{tabular}[c]{@{}l@{}}Distributed\\storage,\\ Tamper-free \\ transactions\end{tabular}                      & \begin{tabular}[c]{@{}l@{}}SMA* gave the optimal\\ path in a very less time\\ and it uses bounded\\ memory as well\end{tabular}                                                                  & \begin{tabular}[c]{@{}l@{}}Efficient dynamic\\ partitioning of the\\ UAV groups is \\ needed\end{tabular}                                \\ \hline
\cite{local}   & \begin{tabular}[c]{@{}l@{}}DoS attacks,\\ Wormhole-\\ attack,\\ GPS spoofing\end{tabular}           & \begin{tabular}[c]{@{}l@{}}The co-ordinates of the drones\\stored in the blockchain is\\made available to the other\\ drones after the verification\end{tabular}                     & \begin{tabular}[c]{@{}l@{}}Decentralized\\ network,\\ Distributed-\\ database\end{tabular}   & \begin{tabular}[c]{@{}l@{}}The localization errors are \\ reduced by $75$\%\end{tabular}                                                                                                         & \begin{tabular}[c]{@{}l@{}}Model is still\\ susceptible to\\ $51$\% attack\end{tabular}                                                  \\ \hline \rowcolor[gray]{0.9}
\cite{BDI2}  & \begin{tabular}[c]{@{}l@{}}Eavesdropping\\ attacks,\\ GPS spoofing,\\ DoS attacks\end{tabular}      & \begin{tabular}[c]{@{}l@{}}If the hash value generated by\\ the Merkle Hash tree and the\\ computed hash value by the\\ forger node is the same then\\ only the data is transmitted\end{tabular}     & \begin{tabular}[c]{@{}l@{}}Data integrity,\\ Distributed- \\ database,\\ Decentralized- \\ network\end{tabular} & \begin{tabular}[c]{@{}l@{}}The total computation\\ time required for data \\ dissemination is computed\\ to be $0.046$ms.\end{tabular}                                                           & \begin{tabular}[c]{@{}l@{}}Implementing\\ private blockchain\\ can make the\\ system more secure\end{tabular}                            \\ \hline
\cite{blotab1} & \begin{tabular}[c]{@{}l@{}}DoS attacks,\\ DDoS attacks,\\ GPS spoofing\end{tabular}                 & \begin{tabular}[c]{@{}l@{}}The data generated by the \\ drone is stored in the \\ blockchain and is transformed\\ into the Merkle Hash tree\\ to maintain the data integrity\end{tabular}       & \begin{tabular}[c]{@{}l@{}}Decentralized-\\ network,\\ Peer-to-Peer\\ model,\\ Immutability\end{tabular}        & \begin{tabular}[c]{@{}l@{}}The results show that only\\ the validated drone\\ were allowed to transfer\\ the data\end{tabular}                                                             & \begin{tabular}[c]{@{}l@{}}The model can be\\ further enhanced\\ for multi UAV\end{tabular}                                              \\ \hline \rowcolor[gray]{0.9}
\cite{blotab2} & \begin{tabular}[c]{@{}l@{}}Man-in-the-\\ middle attack,\\ DoS attack,\\ DDoS attack\end{tabular}    & \begin{tabular}[c]{@{}l@{}}The swarms of drones needs to\\ register themselves on the \\ blockchain using their public\\ key and after the validation,\\ the data is added to the server\end{tabular} & \begin{tabular}[c]{@{}l@{}}Distributed-\\ Database,\\ Immutability,\\ Public Key-\\ infrastructure\end{tabular} & \begin{tabular}[c]{@{}l@{}}The data acquisition was\\ done successfully and with\\ high efficiency\\\end{tabular}                                                                                  & \begin{tabular}[c]{@{}l@{}}Blockchain with\\ higher TPS can be\\ incorporated in the\\ model for making\\ it more efficient\end{tabular} \\ \hline
\cite{blotab3} & \begin{tabular}[c]{@{}l@{}}DoS attacks,\\ DDoS attacks,\\ Man-in-the-\\ middle attacks\end{tabular} & \begin{tabular}[c]{@{}l@{}}The trust is lost from\\ the intruding UAV when\\several intruder events\\are detected\end{tabular}                                 & \begin{tabular}[c]{@{}l@{}}Peer-to-Peer \\ model,\\ Decentralized-\\ network\end{tabular}                       & \begin{tabular}[c]{@{}l@{}}$90$\% of the UAVs were\\ able to support the data\\ about the event and none\\ detected the intruder event\end{tabular}                                              & \begin{tabular}[c]{@{}l@{}}Work on detecting\\ the compromised\\ UAV successfully\\ is required\end{tabular}                             \\ \hline \rowcolor[gray]{0.9}
\cite{blotab4} & \begin{tabular}[c]{@{}l@{}}GPS spoofing,\\ Man-in-the-\\ middle attacks\end{tabular}                & \begin{tabular}[c]{@{}l@{}}A consumer makes an order \\ according to which a smart\\ contract is generated. Any\\ free UAV can accept the order\\ and the client details are sent.\end{tabular}      & \begin{tabular}[c]{@{}l@{}}Peer-to-Peer \\ model,\\ Decentralized-\\ network,\\ Smart Contract\end{tabular}     & \begin{tabular}[c]{@{}l@{}}The blockchain and smart\\ contracts are proved to be \\ successful in organizing a\\ secure communication\\  between the UAVs\end{tabular}                           & \begin{tabular}[c]{@{}l@{}}Implementing\\ private blockchain\\ can make the\\ system more secure\end{tabular}                            \\ \hline
\cite{blotab5} & \begin{tabular}[c]{@{}l@{}}DoS attacks,\\ DDoS attacks,\\ Hijacking\end{tabular}                    & \begin{tabular}[c]{@{}l@{}}The interest-key-content\\ binding (IKCB) is stored in the\\ blockchain which is compared\\ by the router and the poisoned\\ data is discarded\end{tabular}               & \begin{tabular}[c]{@{}l@{}}Tamper-\\resistant\\ledger,\\ Consensus-\\ algorithm\end{tabular}                     & \begin{tabular}[c]{@{}l@{}}The proposed model gives\\ better results when\\ compared with the\\ Interest-key binding as it\\ has lower system overhead\\ and the latency is reduced\end{tabular} & \begin{tabular}[c]{@{}l@{}}The forwarding\\ technologies can\\ be optimized to\\ make the model\\ efficient\end{tabular}                 \\ \hline
\end{tabular}
\label{relblo}
\end{table*}

\subsection{ \textit{\textbf{\textcolor{black}{Secure Localization}}}}
A swarm of drones are deployed that automatically take actions to achieve a specific goal cooperatively. In such scenarios, the exact location coordinates of the drones is critical information for completion of the mission. However, the generic localization algorithms are vulnerable to attacks as the adversary can easily inject false location coordinates. A pioneering work on secure localization on the internet of drones using blockchain is presented in \cite{local}. The authors propose a blockchain-based localization algorithm for securing drones. Three major features of the algorithm are, (i) Decentralization: no central entity would be present to maintain the localization coordinates of the drone, (ii) Peer-to-Peer communication between the drones, and (iii) no need for a central trust node to manage the security of the data and the coordinates between the drones. The other non-blockchain based approach for securing localization in IoD is discussed in \cite{worm}. This method focuses only on preventing the wormhole attack. The approach in \cite{worm} being centralized in nature cannot be used to prevent other generic attacks such as DoS. In the proposed algorithm, the drones need to cooperate with various other anchor drones knowing their exact coordinates. The coordinates of the anchors are sent to the requesting drone using the private key of the anchors. Next, the coordinates of the anchor drones are added to the distributed blockchain ledger after verification. The requesting drone first requests for the coordinates of the anchor drones present at a $1$-hop distance. If the requesting drone receives the location from at least three anchor drones in the 1-hop, the distance between the requesting drone and the anchor drone is calculated using the Received Signal Strength Indicator (RSSI)  method \cite{RSSI}. However, if the requesting drone does not receive three minimum responses from the neighboring anchor drones, the distance between the requesting drone and the anchor drones is calculated using the DV-Hop (Distance Vector Hop) method \cite{DVHOP}. The DV-Hop method works by first computing the average hop distance and then multiplying it by the number of hops. The authors also calculate the change in localization accuracy with the increase in the number of malicious nodes in the network. The accuracy of the proposed algorithm is proven to be better than the generic localization algorithms \cite{genloc}. Simulation results demonstrate that the localization errors are minimized by $1/4^{th}$ in the presence of $50$\% malicious nodes in the network. 
Moreover, due to the decentralized nature of blockchain, various other attacks such as DoS attack, and wormhole attack can also be prevented. However, the model is still susceptible to the $51$\% attack and other attacks, as is the case with blockchain. $51$\% attack happens when the number of malicious nodes in the network become more than half of the total nodes in the network, and hence, fair localization coordinates would not be revealed.

\par 

\textbf{       Summary:} 
\textcolor{black}{The objective here is to minimize the possibilities of any kind of physical attack on drones or data losses in drone communication. A summary of advantages and disadvantages of major works is given in Table \ref{bloadv}, and a summary of the related works is given in Table \ref{relblo}. As seen, the blockchain technology is mostly used to provide a peer-to-peer model to mitigate the various security issues related to generic centralized architectures. In various works, the smart contract and incentive model features of blockchain are also used to enhance data security and reliability in various scenarios related to drone communication.} 


\section{Applications of SDN for Drone communication Security}
\label{sec6}Software-Defined Networking may be defined as a networking paradigm centered around the separation of the data plane or forwarding plane of a computer network from its control plane, and the application layer. Software-defined networking is an architecture that aims to make networks agile and flexible. The SDN virtualizes the network by separating the control plane that manages the network from the data plane where all the traffic flows. SDN simply decouples the network control from the forwarding process of the network packets. This decoupling allows the network to be controlled separately without worrying about the traffic flow. This infrastructure will keep the traffic and the network services abstracted from the network control. These SDN parameters help in making the drone communication secure. In this section, we talk about the SDN-based DCN that helps in resolving various security issues in DCN.

There are several existing solutions that tend to resolve the security issues related to drone communication without using the SDN architecture. The pioneer work in this direction is presented in \cite{non2}. In this model, the UAV receives the request from the ground controller and sends the data back to the controller in the form of visuals. The method described in the model demands high bandwidth for its execution, which varies with the speed of the drone or the broadcasting channels. Such traditional solutions fail to provide high security in the new generation drones, and also fail in maintaining the data integrity. Another model proposed in \cite{non3} uses heuristic algorithms for providing data integrity but fails in providing good efficiency and reliability.

\begin{table*}[]
\centering
\caption{A summary of advantages and disadvantages of major applications of SDN for drone communication security.}
\begin{tabular}{|l|l|l|l|}
\hline \rowcolor[gray]{0.8}
\begin{tabular}[c]{@{}l@{}}Major\\ approaches\end{tabular} & Advantages & Disadvantages  & \begin{tabular}[c]{@{}l@{}} \textcolor{black}{Benefits over}\\ \textcolor{black}{traditional approaches}  \end{tabular}       \\
\hline
\cite{16}                                 & \begin{tabular}[c]{@{}l@{}}• Very short average file-transfer time\\ • Every sender gets fair share of\\ bandwidth\end{tabular}   & \begin{tabular}[c]{@{}l@{}}• Legitimate sender may have to\\ wait for a long time\end{tabular}    & \begin{tabular}[c]{@{}l@{}} \textcolor{black}{Scalability, Lightweight,}\\ \textcolor{black}{Self-reliant defense} \end{tabular}                                \\
\hline
\cite{DS1}                              & \begin{tabular}[c]{@{}l@{}}• The drone from which the DDoS\\ attack is launched can be found in\\ a very less time\end{tabular}                                                  & \begin{tabular}[c]{@{}l@{}}• Cannot work properly when the\\ number of flow table items in the SD-\\IoT is very high\end{tabular}  & \begin{tabular}[c]{@{}l@{}} \textcolor{black}{Detecting and Mitigating}\\ \textcolor{black}{DDoS attacks faster} \end{tabular}     \\
\hline
\cite{intent}                             & \begin{tabular}[c]{@{}l@{}}• The average end-to-end outage rate\\ in IoD is reduced by $18$\%\end{tabular}                                                                          & • High average end-to-end delay     &   \begin{tabular}[c]{@{}l@{}} \textcolor{black}{Reduction in end-to-end}\\ \textcolor{black}{delay} \end{tabular}   \\
\hline
\cite{sdn1}                               & \begin{tabular}[c]{@{}l@{}}• The model has high fault-tolerance\\  • High performance due to the presence\\ of multiple controllers\end{tabular} & \begin{tabular}[c]{@{}l@{}}• The link between the data plane and\\ the control plane is still susceptible\\ to attacks\end{tabular} & \begin{tabular}[c]{@{}l@{}} \textcolor{black}{Scalability, Mobility} \end{tabular}\\
\hline
\cite{SI1}                              & \begin{tabular}[c]{@{}l@{}}• The latency and the maximum load\\ experienced by a SDN switch is reduced\\ by $50$\%\end{tabular}     & \begin{tabular}[c]{@{}l@{}}• The complexity of the algorithm is\\ quite high \end{tabular} & \begin{tabular}[c]{@{}l@{}} \textcolor{black}{Stability, Security,}\\ \textcolor{black}{Reduced network latency} \end{tabular}\\ \hline                                        
\end{tabular}
\label{sdnadv}
\end{table*}
Considering the above issues, in this section, we discuss the methods that involve SDN for maintaining the security in drone communication. We  further  discuss  the  list  of  specific  security  issues  that can be resolved and prevented using SDN as a solution.

\subsection{\textit{\textbf{\textcolor{black}{DoS Attacks}}}}
Due to resource constraints, we need a highly efficient protocol that is resistant to large–scale DoS attacks. NetFence protocol in SDN based Drone Communication Network (DCN) as proposed in \cite{16} can be used to create a scalable DoS resistant network. The proposed model makes use of traffic policing inside the network. The packets in the network carry the unforgeable congestion policing feedback that is attached on the packets by routers. For a drone to be a part of the network, it needs to first send a request packet to a NetFence ready receiver. When it is accepted, it receives feedback and along with the acknowledgement, it sends regular data packets. Non-NetFence senders can only send packets through the legacy channel which is given the lowest packet-forwarding priority. Bottleneck routers act as congestion detectors which regularly check link load and packet loss rate. The rate limiters reduce data congestion through the Additive Increase and Multiplicative Decrease (AIMD) algorithm \cite{AIMD}. NS-$2$ simulations were implemented in linux, and the performance of NetFence in DoS attacks was compared with $3$ other mechanisms that are Traffic Validation Architecture (TVA+) \cite{TVA}, StopIt \cite{stop}, and Fair Queuing (FQ) \cite{FQ}. Netfence has the advantage of having short average file-transfer time that does not increase significantly with an increase in senders, whereas, in FQ, transfer time increases linearly with an increase in senders. Even though mechanisms like TVA+ and StopIt tend to block large-scale DoS attacks, per-host queuing is implemented in these algorithms as compared to per-Autonomous System queuing in NetFence. This is advantageous as the number of autonomous systems is significantly less than the number of hosts. No matter how heavy the attack is, the Netfence protocol makes sure that senders get their fair share of bandwidth. This model has a drawback, a legitimate sender may need to wait more time in Netfence to transmit data than in TVA+, StopIt. Additionally, NetFence algorithm also fails to distinguish between the congestion caused by DoS attack or any other issue in the network.

\subsection{\textit{\textbf{\textcolor{black}{Distributed Denial of Service (DDoS) attacks}}}}
The DDoS attack is a bit different from a normal DoS attack. In a DDoS attack, there are multiple numbers of compromised hosts as compared to a single compromised host in a normal DoS attack. An intelligent and lightweight approach is required to prevent DDoS attacks in IoD. A pioneer work in avoiding DDoS attacks in IoT using SDN is \cite{DS1}. Unlike \cite{16}, the authors of \cite{DS1} propose an algorithm to detect and mitigate the DDoS attacks in drones. Cosine similarity of the vectors of the packet-in message rate at software-defined Internet of Things (SD-IoT) switch ports is used to determine the occurrence of DDoS attack. The threshold values of the vector length and cosine similarity are used to precisely and accurately classify an attack situation. The simulation results demonstrate that the proposed algorithm is capable of detecting the device used to launch the DDoS attack in a short span of time. The results of the proposed work are compared with other state-of-the-art work that try to detect DDoS attacks using IP filtering \cite{DDoS2}. The simulation results demonstrate that in case of a DDoS attack, the number of flow table items of the SD-IoT switches, and the number of data packets received by the SD-IoT controller are less in \cite{DS1} as compared to \cite{DDoS2}. However, a proactive scheme to defend and prevent the DDoS attacks is missing. The proposed algorithm will work only after the DDoS attack has been launched. The authors of \cite{conti3} provide a lightweight solution to counter DDoS attacks using SDN.

\subsection{\textit{\textbf{\textcolor{black}{Avoiding intentional disruption}}}}
Apart from DDoS attacks on a set of drones in the network, the network of drones being resource-constrained is also susceptible to intentional jamming and disruption attacks. Such attacks are more severe as compared to DDoS attacks, as they can paralyze the entire network leaving no room for detection and mitigation. Different from \cite{16} and \cite{DS1}, the authors of \cite{intent} have proposed an SDN-based framework for secure and robust data relaying in case of intentional jamming and disruptions. In the proposed model, the drones act as SDN switches that are controlled by a centralized SDN controller. A novel 3D spatial coverage metric is used to calculate diverse multiple paths among the drones to prevent the effect of intentional disruptions on the functioning of the drone network, as the model gives the directives to the drone for using the best possible path. The simulation results demonstrate that the proposed algorithm outperforms the traditional shortest path and shortest multi-path algorithms in terms of outage prevention \cite{intentional2}. The average end-to-end outage rate in IoD is reduced by $18$\% in the proposed model when compared to \cite{intentional2}. The algorithms in \cite{intentional2} consider only the path that takes the least time, irrespective of the presence of jammers and intentional disrupts. Although the proposed model in \cite{intent} succeeds in preventing the complete outage of the drone network, the end-to-end delay is also increased by $12$\% when compared to \cite{intentional2}. The proposed model in \cite{intent} also helps in preventing the frequent link disconnections between the devices linked to the network. Further work is required to propose some algorithms that can prevent such intentional disruptions without increasing the average delay of the network. The jamming incident in Hong Kong drones could've been prevented if any of the above-mentioned measures were taken \cite{jammerinc}.

\begin{table*}[]
\centering
\caption{Applications of SDN for drone communication security. }
\begin{tabular}{|p{0.04\linewidth}|p{0.075\linewidth}|p{0.23\linewidth}|p{0.125\linewidth}|p{0.2\linewidth}|p{0.18\linewidth}|}
\hline \rowcolor[gray]{0.7}
Ref.                           & Attack                                                                                                 & Mechanism                                                                                                                                                                                                           & \begin{tabular}[c]{@{}l@{}}SDN Feature \\ used\end{tabular}                                                           & Major achievement                                                                                                                                                     & Open issues                                                                                                                                                            \\
\hline
\cite{16}     & DoS attacks                                                                                            & \begin{tabular}[c]{@{}l@{}}The drone first registers\\ itself with the NetFence\\ ready receiver and then\\ only the drone is allowed\\ to transmit the data packets.\end{tabular}                                  & \begin{tabular}[c]{@{}l@{}}Directly-\\ programmable,\\ Scalability\end{tabular}                                       & \begin{tabular}[c]{@{}l@{}}The model has a very\\ short average file-transfer\\ time\end{tabular}                                                                     & \begin{tabular}[c]{@{}l@{}}Implementing the\\ model specifically\\ for UAVs is much\\ needed\end{tabular}                                                              \\
\hline \rowcolor[gray]{0.9}
\cite{DS1}    & \begin{tabular}[c]{@{}l@{}}DDoS\\ attacks\end{tabular}                                                 & \begin{tabular}[c]{@{}l@{}}The cosine similarity of the\\ vectors of the packet-in\\ message rate at the SD-IoT\\ switch port is used to\\ determine the attack\end{tabular}                                        & \begin{tabular}[c]{@{}l@{}}Abstraction of\\ network devices,\\ Dynamic re-\\ configuration of\\ networks\end{tabular} & \begin{tabular}[c]{@{}l@{}}Can determine the device\\ using which the DDoS\\ attack is launched in a\\ very short time span\end{tabular}                              & \begin{tabular}[c]{@{}l@{}}Implementing the\\ model specifically\\ for UAVs is much\\ needed\end{tabular}                                                              \\
\hline
\cite{intent} & \begin{tabular}[c]{@{}l@{}}Jamming\\ attacks,\\ Disruption\\ attacks\end{tabular}                      & \begin{tabular}[c]{@{}l@{}}Multiple paths are generated\\ using a $3$D spatial metric\\ which are directed to the\\ UAVs to avoid the disruption\end{tabular}                                                       & \begin{tabular}[c]{@{}l@{}}Decoupled Data-\\ plane and the\\ Control plane\end{tabular}                               & \begin{tabular}[c]{@{}l@{}}The average end-to-end\\ outage rate in the IoD is\\ reduced to a large extent\\\end{tabular} & \begin{tabular}[c]{@{}l@{}}End-to-end delay\\ increases significantly,\\ which is a major area\\ to be looked upon in\\ the future\end{tabular}                        \\
\hline \rowcolor[gray]{0.9}
\cite{sdn1}   & \begin{tabular}[c]{@{}l@{}}DoS\\attacks,\\ GPS\\ Spoofing\end{tabular}                                  & \begin{tabular}[c]{@{}l@{}}SDN controller authenticates\\ the network device and then\\ only the data is transmitted\\ by the controllers\end{tabular}                                                              & \begin{tabular}[c]{@{}l@{}}Decoupled Data-\\ plane and the\\ Control plane\end{tabular}                               & \begin{tabular}[c]{@{}l@{}}Multiple SDN controllers\\ are deployed to prevent\\ the malfunctioning of the\\ devices in the network\end{tabular}                       & \begin{tabular}[c]{@{}l@{}}The link between the\\ Control plane and Data\\ plane is still\\ susceptible to attacks\end{tabular}                                        \\
\hline
\cite{SI1}    & \begin{tabular}[c]{@{}l@{}}DoS\\attacks,\\ Spoofing\\ attacks\end{tabular}                              & \begin{tabular}[c]{@{}l@{}}The Middlebox-Guard (M-G)\\ is deployed at different\\ locations which manages the\\ dataflow\end{tabular}                                                                               & \begin{tabular}[c]{@{}l@{}}Directly-\\ programmable,\\ Flexible network\\ architecture\end{tabular}                   & \begin{tabular}[c]{@{}l@{}}Latency and the\\ maximum load on the\\ device are reduced by\\ $50$\%\end{tabular}                                                        & \begin{tabular}[c]{@{}l@{}}The Integer Linear\\ Program (ILP) pruning\\ algorithm used in M-G\\ has a high complexity\end{tabular}                                     \\
\hline \rowcolor[gray]{0.9}
\cite{sdnt1}  & \begin{tabular}[c]{@{}l@{}}DoS\\attacks,\\ DDoS\\ attacks\end{tabular}                                  & \begin{tabular}[c]{@{}l@{}}The primary path forwards\\the common files whereas\\the backup path forwards\\the uncommon cases where\\the primary path is not\\reliable\end{tabular}                              & \begin{tabular}[c]{@{}l@{}}Decoupled Data-\\ plane and the \\ Control plane,\\ Scalability\end{tabular}               & \begin{tabular}[c]{@{}l@{}}It can handles link\\ congestion with high\\ bandwidth\end{tabular}                                                                        & \begin{tabular}[c]{@{}l@{}}It has a very high end-\\ to-end delay, which\\ has to be looked upon\\ in the future to make\\ the algorithm more\\ efficient\end{tabular} \\
\hline
\cite{sdnt2}  & \begin{tabular}[c]{@{}l@{}}DoS\\attacks,\\ DDoS\\ attacks\end{tabular}                                  & \begin{tabular}[c]{@{}l@{}}SDN computes the optimal\\ flow for each multi-path TCP\\ and the Flow Deviation\\ Method (FDM) algorithm\\ is used to re-allocate\\ the bandwidth\end{tabular}                          & \begin{tabular}[c]{@{}l@{}}Network-\\ programmability,\\ Decoupled Data-\\ plane and the\\ Control plane\end{tabular} & \begin{tabular}[c]{@{}l@{}}The model achieves\\ fairer bandwidth\\ allocation that provides\\ better QoS and it makes\\ the network more reliable\end{tabular}        & \begin{tabular}[c]{@{}l@{}}Cannot support a high\\ number of users and\\ the model is not fully\\ secure\end{tabular}                                                  \\
\hline \rowcolor[gray]{0.9}
\cite{fig11}  & \begin{tabular}[c]{@{}l@{}}Grey hole\\attacks,\\Black hole\\ attacks,\\DDoS\\attacks\end{tabular} & \begin{tabular}[c]{@{}l@{}}The UAV informs its\\ controller about the\\ neighboring drone while\\ establishing OpenFlow\\ connection and also informs\\ about its update\end{tabular}                               & \begin{tabular}[c]{@{}l@{}}Decopuled Data-\\ plane and the\\ Contol plane,\\ Scalability\end{tabular}                 & \begin{tabular}[c]{@{}l@{}}The amount of data\\ exchange when compared\\ with the AODV routing\\ algorithm is increased by\\ $2$\%\end{tabular}                     & \begin{tabular}[c]{@{}l@{}}Further works on\\ increasing the security\\ of the model is needed\end{tabular}                                                            \\
\hline
\cite{sdnt3}  & \begin{tabular}[c]{@{}l@{}}GPS\\ Spoofing\end{tabular}                                                 & \begin{tabular}[c]{@{}l@{}}Cluster heads are assigned to\\different densely populated\\sectors and the data is\\transferred through the cluster\\head only when in the range\end{tabular} & \begin{tabular}[c]{@{}l@{}}Flexible network\\ architecture,\\ Scalability\end{tabular}                                & \begin{tabular}[c]{@{}l@{}}The model provides faster\\ and efficient coverage rate\\ of about $99$\% and a\\ latency of around $20$\%\end{tabular}                    & \begin{tabular}[c]{@{}l@{}}Need to decrease\\ the latency to make\\ the model more\\ efficient\end{tabular} \\    \hline                                                      
\end{tabular}
\label{relsdn}
\end{table*}

\subsection{ \textit{\textbf{\textcolor{black}{Malfunctioning devices}}}}
Apart from DoS, DDoS, and intentional jamming, there are various other issues in drone communication that are related to the different sensors deployed on the drones. Traditional internet systems use IP and firewalls, which cannot solve these issues, as it is not possible to fit all objects and protocols to a common and singular protocol. A lightweight model for avoiding malfunctioning devices in IoD is proposed in \cite{sdn1}. In the proposed model, the SDN controller first authenticates the network device requesting to be connected to the network. Only after successful authentication, the data is disseminated to the connected devices using the controller that makes sure that no malfunctioning of the device is taking place. Traditional network protocols are not designed to support high levels of traffic, scalability, and mobility. Hence, the use of SDN in this work increases the functionality of the network by reducing the hardware complexity. SDN also has the ability to extend the network security to the access end-point devices. Multiple SDN controllers have been used instead of a single one to improve fault tolerance and robustness. Unlike \cite{sdn1}, the authors of \cite{sdn2} have proposed a similar framework using only a single controller. If an attacker compromises the SDN controller, he gains full control over the network. Hardware and software failures may also occur, which pose a potential risk to the entire network. The work in \cite{sdn1} is superior as it uses multiple controllers, so if one goes down, another can take control to avoid a system failure in case of any malfunctioning device. The proposed work reports increased network performance with multiple controllers because each controller has a partial view of the network and the controllers collaborate and exchange information with each other. However, the link between the Control Plane and Data Plane of the SDN is still vulnerable and susceptible to attacks and these issues are yet to be resolved.

\subsection{ \textit{\textbf{\textcolor{black}{Data Integrity}}}}
An SDN-based data security model Middlebox-Guard (M-G) is proposed in \cite{SI1}. Different from \cite{sdn1}, and \cite{sdn2}, the M-G manages the dataflow to ensure network efficiency and helps in minimizing the network latency. To reduce the latency, the middleboxes are placed at locations where the communication link is the shortest using a placement selection algorithm. The middleboxes are placed in different locations and an offline Integer Linear Program (ILP) pruning algorithm \cite{ILP} is deployed at each middlebox. ILP helps in solving uncontrollable computation optimization problems at every middlebox to tackle the switch constraints, such as the use of CPU, RAM, etc. The ILP algorithm also provides the optimum routes to be used for the data transfer. Also, an online ILP is used to minimize the maximum middlebox load across the network. M-G is compared to a model known as SIMPLE proposed in \cite{SDNI2}, as both solve middlebox placement and route selection problems. M-G outperforms the latter in terms of security, latency, and load. In \cite{SI1}, POX was used as the controller, and OpenvSwitch was used as the SDN switch for carrying out the experiments. On running the entire system, latency and maximum loads were reduced by $50$\%. In terms of security, middlebox failures and overload conditions were analyzed, and the response times for these were calculated to be less than $0.15$ seconds, which means, it shows a fast response.
 
 \par
 \textbf{       Summary: }
\textcolor{black}{SDN can help in preventing many attacks that drones are susceptible to, including DoS attacks and DDoS attacks, and can help in maintaining data integrity in drones. SDN technology can also help in avoiding the intentional disruption and jamming attacks that impose danger on drone communication. A summary of the advantages and disadvantages of major works that use SDN as a solution to drone communication security are described in Table \ref{sdnadv}. Furthermore, Table \ref{relsdn} summarizes the related works that use SDN in maintaining security in drone communication. According to the best of our knowledge, the decoupling of the control plane, the data plane, and the network plane helps a lot in maintaining security standards in drone communication.} 

\section{Applications of Machine Learning for Drone Communication Security}
\label{sec7}Machine learning is the study of algorithms that are capable of learning and improving automatically through experience, and can make accurate predictions based on the data with which they are fed \cite{ml}. They can provide generalized observations for unseen and unknown states and networks as well. Different machine learning algorithms are useful in different drone applications and domains. The use of specific ML algorithms is dependent on the domain, and type of data available. The ML approaches have been extensively explored in literature both for physical security of drones, and for drone communication security. The physical security approaches basically deal with using different ML algorithms to detect unauthorized drones or to prevent authorized drones from entering into unauthorized zones. Both these types of security issues are intrinsically related to each other. For example, if the system fails to identify or detect an unauthorized drone and allows it to enter into a network of authorized drones, it can easily allow all possible communication attacks on the network. Therefore, drone detection using ML can be considered as a preliminary step that can prevent the possibility of drone communication issues to a great extent. The authors of \cite{jam_ml} study different ML frameworks and provide a model to prevent jamming attacks. A distributed learning framework is essential to manage the various tasks in a swarm of drones \cite{mozaffari5}. Therefore, in this section, we review the various works that try to use various ML algorithms to detect the drones or to identify and prevent the generic security vulnerabilities. First of all, we discuss the issues with the traditional approaches that do not use ML algorithms, and then we move on to the challenges in drone communication and possible ML-based solutions. 

There are some traditional techniques that do not use ML algorithms for detecting drones. The most primitive technique is drone detection using radar. Detection using radar is highly expensive, and it can be used only for detecting large objects. Another model that can be used for drone detection is using Light Detection And Ranging (LiDAR) that has been implemented in \cite{LiDAR}. LiDAR sends the laser beam towards the object and analyzes the beams returned after colliding with the object. However, LiDAR is also a extremely expensive method for detection, and is highly vulnerable to climatic conditions. Moreover, these techniques tend to give false positive results, thereby resulting in wastage of resources. 

We  further discuss specific security issues that can be resolved and prevented using ML as a solution.
\subsection{\textit{\textbf{\textcolor{black}{Drone detection using SVM and CNN}}}}
ML algorithms can be used in radar detection to address various detection and classification problems associated with the traditional methods of radar detection \cite{radar}. The authors discussed different SVM models to classify the detected objects as drones or birds, classify different kinds of drones depending on payload or number of rotors. These models showed high accuracy on test data (>$90$\%). An efficient drone detection model using the Support Vector Machine (SVM) and Convolutional Neural Networks (CNN) for noise detection is discussed in \cite{SVM}. Unlike \cite{ML1}, the authors of \cite{SVM} use SVM and CNN for drone detection as compared to LSTM approach used in \cite{ML1}. \textcolor{black}{The data was collected using audio from $6$ listening nodes to listen to the UAV flown \cite{SVM}.} Both types of ML algorithms have their own pros and cons. The SVM-based models are easier to implement as compared to other deep learning algorithms such as LSTM. However, the SVM-based models are only suitable for small datasets with limited outliers. In the proposed model, multiple listening nodes and a control center are used. The listening nodes are deployed on a circle surrounding the protected area. A microphone is installed on the listening nodes, to detect the sound of the drone. After the detection, the modules per frame are computed, and are sent to the control center for further evaluation. At the control center, SVM is deployed. SVM is a supervised classifier that helps in classifying between the required entities by mapping the input vectors into a high-dimensional feature space \cite{SVM2}. This classifier plots the pattern of the frames and the sound that is sent to the control center, and plots whether the drone is detected or not. \textcolor{black}{The simulation results in \cite{SVM} demonstrate that the SVM algorithm is more efficient than CNN in detecting drones.} However, the main limitation is that these algorithms have noise-related issues that make the results inconsistent. Moreover, the signals were not normalized in the proposed model leading to a lot of outliers.

\begin{table*}[]
\centering
\caption{A summary of advantages and disadvantages of major applications of ML for drone communication security.}
\begin{tabular}{|l|l|l|l|}
\hline \rowcolor[gray]{0.8}
\begin{tabular}[c]{@{}l@{}}Major   \\ Approaches\end{tabular} & Advantages   & Disadvantages & \begin{tabular}[c]{@{}l@{}} \textcolor{black}{Benefits over}\\ \textcolor{black}{traditional approaches} \end{tabular}  \\ 
\hline

\cite{SVM}, \cite{CNN}    & \begin{tabular}[c]{@{}l@{}}• Supports high computation speed\\ and is cost efficient\end{tabular}                  & \begin{tabular}[c]{@{}l@{}}• Unwanted noises in   the background\\ makes the results inconsistent\end{tabular} & \textcolor{black}{Efficiency}        \\ \hline
\cite{RNN}, \cite{RNNIMAGE}                                 & \begin{tabular}[c]{@{}l@{}}• Model training time is very less\\ and gives high accuracy\end{tabular}                & \begin{tabular}[c]{@{}l@{}}• Generating such a large dataset\\ artificially is very difficult\end{tabular}  & \begin{tabular}[c]{@{}l@{}} \textcolor{black}{Identification, Classific-}\\ \textcolor{black}{ation of types of drones} \end{tabular}                     \\ \hline
\cite{inter}                               & \begin{tabular}[c]{@{}l@{}}• Stores data sequentially, so the data\\ retrieval latency is very less\end{tabular}    & \begin{tabular}[c]{@{}l@{}}• The latency in data transmission\\ increases when a very large data file\\ is transmitted\end{tabular} & \begin{tabular}[c]{@{}l@{}} \textcolor{black}{Minimizing latency, }\\ \textcolor{black}{Data reliability} \end{tabular} \\ \hline
\cite{neural}                              & \begin{tabular}[c]{@{}l@{}}• High reliability with very fewer\\ resource requirements\end{tabular}                  & \begin{tabular}[c]{@{}l@{}} • Fails in further classifying the type\\ of attack \end{tabular} & \begin{tabular}[c]{@{}l@{}} \textcolor{black}{Detecting and preventing}\\ \textcolor{black}{DoS attacks}  \end{tabular} \\ \hline
\cite{obs}                                 & \begin{tabular}[c]{@{}l@{}}• Very lightweight model, can run\\ on Raspberry Pi $3$B\end{tabular}                   & \begin{tabular}[c]{@{}l@{}}• A lot of training and testing in Deep \\ Learning algorithms may be required\end{tabular} & \begin{tabular}[c]{@{}l@{}} \textcolor{black}{Easy deployment,}\\ \textcolor{black}{Privacy} \end{tabular}           \\ \hline
\cite{GPSD}                           & \begin{tabular}[c]{@{}l@{}}• Can detect the adversary in GPS-\\ denied environment with great\\ accuracy\end{tabular} & \begin{tabular}[c]{@{}l@{}}• Fails when the drone moves in an\\ irregular pattern and is sensitive to\\ lighting conditions\end{tabular} & \begin{tabular}[c]{@{}l@{}} \textcolor{black}{Efficiency in GPS-}\\ \textcolor{black}{denied environment} \end{tabular}\\ \hline
\end{tabular}
\label{mladv}
\end{table*}
\subsection{\textit{\textbf{\textcolor{black}{Drone detection using RNN and CRNN}}}}
An efficient model using deep learning techniques like Recurrent Neural Networks (RNN) and Convolutional Recurrent Neural Networks (CRNN) for drone detection is mentioned in \cite{RNN}. The authors of \cite{RNN} have acquired a large dataset of drone propeller audio data, and have overlapped the audio clips with a variety of background noises to mimic real-life scenarios. \textcolor{black}{Data labelling was done for the identification problem as unknown (random noises in the surrounding), Bebop (drone $1$), or Mambo (drone $2$), and for the detection problem as drone or not a drone.} The experiment has been divided into two categories, the first, targeting detection of drones and the second, targeting their identification based on type. The detection problem has been evaluated and compared with existing literature, and the mentioned algorithms have been compared based on their accuracy, F1 score, precision, recall metrics, and computational time required to train and test the model. The \textcolor{black}{experiment} results of the model in \cite{RNN} show that deep learning methods using drone acoustics show great effectiveness for drone detection. CNN and CRNN algorithms remarkably outperform RNN in both detection and identification. Although CNN showed better performance than CRNN, the difference in performance was negligible, and CRNN required significantly less time to train, making it the most practical choice. Another model as discussed in \cite{RNNIMAGE} uses CNN for drone detection, but uses images instead of drone acoustics. Although the results are promising, the dataset for such a study could only be artificially created which decreases its reliability, and identification of the specific type of drone is also not possible like it is in \cite{RNN}. The authors of \cite{RNNWIN} have noted that RNN achieves superior performance when compared to CNN. The discrepancy is attributed to differences in the model’s architecture and design parameters, but a direct comparison of the results could not be performed by the authors of \cite{RNN}.

\subsection{\textit{\textbf{\textcolor{black}{Fault detection and Recovery of UAV data using LSTM}}}}
UAVs are used for certain critical applications like military, and product delivery. Therefore, it is imperative to deploy a certain mechanism for making data transmission in UAVs ultra-reliable. Furthermore, the latency of data transmission should also be kept minimum. Being resource-constrained, the UAVs need to transmit real-time data to the cloud servers for storing. Pioneer works in the direction of minimizing the latency and increasing the data reliability using LSTM are \cite{ML1}, \cite{inter}. Unlike \cite{RNN}, the authors of \cite{ML1}, \cite{inter} use LSTM for drone communication security. LSTM networks are a special type of RNN networks with some special features. The main feature of LSTM over RNN is the presence of a 'memory cell' that can maintain information in memory for a long time. In the proposed model, firstly, a regression model using LSTM is built to extract spatial-temporal features of the drone data. This is done to get an estimate of the monitored parameters or features. The authors use a set of $11$ distinct parameters or features, like roll angle, altitude, indicated airspeed, etc., to sense the UAV's current attitude and position through airborne sensors as an input to the proposed model. The output is used to train the fault detection model after the normalization of the data. Next, various filters are used to reduce the difference between the actual data and estimated values, thereby removing the effects of various noises. A threshold value is compared with the estimated values to detect the faults. In case a fault in data is discovered, the faulty data is replaced with the data estimated by the proposed model or the recovery data. \textcolor{black}{The simulation results demonstrate that the proposed model is capable of providing a quick recovery of the data in a limited time. The experimentation shows that the Mean Square Error (MSE) was recorded to be less than $0.078$ whereas, the Mean Absolute Error (MAE) was less than $0.205$.} However, further work on increasing real-time data recovery may be done to make the model more accurate and effective in fault detection and recovery.

\subsection{\textit{\textbf{\textcolor{black}{DoS attacks}}}}
The authors of \cite{dist}, and \cite{msvm} have proposed an ML-based model to detect Denial of Service Attack using Neural Networks and Modified Support Vector Machines respectively. However, a pioneer model for detecting and preventing DoS attacks in IoD using machine learning is proposed in \cite{neural}. Unlike \cite{SVM,RNN,inter}, the authors of \cite{neural} focus on preventing the DoS attacks on drone data, rather than physically detecting the unwanted drones in the network. \textcolor{black}{The dataset consists of labeled data categorized as Benign for normal traffic, and attacks like brute force, DoS/DDoS, and web attacks.} The authors proposed and implemented the random forest algorithm \cite{RF} and multi-layer perceptron algorithm \cite{MLP} on the CIC IDS $2017$ dataset. CIC IDS $2017$ dataset consists of all the data of current attacks, such as DoS and DDoS, in pcap format. The incoming data traffic in the drone is classified using the deployed classification algorithms to be as benign or affected packet. In both of the models, an accuracy greater than $98$\% was achieved with the MLP achieving an accuracy of $98.87$\% with $30$\% training records and the RF algorithm achieving an accuracy of $99.95$\% with $50$\% training records. However, none of the previous works including \cite{dist} and \cite{msvm} could achieve such an accuracy level with a relatively low resource requirement, as desired by an IoD system. The further task is to test the system for the multi-classification of DoS attacks. Moreover, the model does not further classify into attacks such as Hearbleed, slowhttptest, and http flood. Also, the resources required can be further reduced to make the system more efficient by reducing the number of features.

\begin{table*}[]
\centering
\caption{Applications of ML for drone communication security.}
\begin{tabular}{|p{0.035\linewidth}|p{0.065\linewidth}|p{0.2\linewidth}|p{0.21\linewidth}|p{0.171\linewidth}|p{0.176\linewidth}|}
\hline \rowcolor[gray]{0.7}
Ref.                                                                                             & Attack                                                                                          & Mechanism                                                                                                                                                                                                                   & \begin{tabular}[c]{@{}l@{}}Machine Learning\\ Feature used\end{tabular}                                                                                                                            & Major achievement                                                                                                                                                  & Open issues                                                                                                                                                                                     \\ \hline
\cite{SVM}                                                                           & \begin{tabular}[c]{@{}l@{}}GPS\\ spoofing\end{tabular}                                          & \begin{tabular}[c]{@{}l@{}}The sound of the drone is \\ used for classifying the\\ presence of drone using\\ SVM and CNN\end{tabular}                                                                                       & \begin{tabular}[c]{@{}l@{}}SVM and CNN \\ classifies whether the\\ drone is present in the\\ specified area or not\end{tabular}                                                                    & \begin{tabular}[c]{@{}l@{}}SVM shows better\\results in detecting\\the UAV as compared\\to CNN\end{tabular}                                                      & \begin{tabular}[c]{@{}l@{}}The background noises\\ of the wind and the\\ surroundings gave\\ inconsistent results\end{tabular}                                                                  \\ \hline \rowcolor[gray]{0.9}
\cite{RNN}                                                                           & \begin{tabular}[c]{@{}l@{}}GPS\\ spoofing\end{tabular}                                          & \begin{tabular}[c]{@{}l@{}}The algorithms like RNN\\ are used to identify\\ the presence of UAV on\\ the basis of sound\end{tabular}                                                                  & \begin{tabular}[c]{@{}l@{}}Algorithms like RNN and\\ CRNN is used to classify\\ the presence of the drone\end{tabular}                                                                             & \begin{tabular}[c]{@{}l@{}}CRNN showed the\\best results in\\detecting the presence\\of the drone\end{tabular}                                                    & \begin{tabular}[c]{@{}l@{}}The model can be\\trained to detect\\ the wide class of\\drones\end{tabular}                                        
\\ \hline
\begin{tabular}[c]{@{}l@{}}
\cite{ML1} \end{tabular} & \begin{tabular}[c]{@{}l@{}}DoS\\ attack,\\ Worm-\\hole\\ attack\end{tabular}                       & \begin{tabular}[c]{@{}l@{}}The LSTM-based fault\\ detection model detects\\ the fault and the quick\\ recovery commands are\\ sent to the UAV\end{tabular}                                                                  & \begin{tabular}[c]{@{}l@{}}LSTM is used to store the\\ previous data of the UAV\\ which helps in building\\ the model that efficiently\\ detects the fault\end{tabular}                            & \begin{tabular}[c]{@{}l@{}}The model achieved\\very less MSE and\\MAE, which makes \\the model very\\efficient\end{tabular}    & \begin{tabular}[c]{@{}l@{}}Work on increasing the\\ efficiency of the model\\ is much needed\end{tabular}                                                                                       \\ \hline \rowcolor[gray]{0.9}
\cite{neural}                                                                        & \begin{tabular}[c]{@{}l@{}}DoS \\ attack\end{tabular}                                           & \begin{tabular}[c]{@{}l@{}}The Random Forest and\\ Multi-Layer Perceptron\\ algorithm classifies\\ the data packets received \\ as benign or the DoS\\ affected packets\end{tabular}                           & \begin{tabular}[c]{@{}l@{}}Random Forest and Multi-\\ Layer perceptron algorithm\\ is used to classify between\\ the affected and the non-\\ affected packet received by\\ the drones\end{tabular} & \begin{tabular}[c]{@{}l@{}}The MLP algorithm\\ achieved an accuracy\\ of $98.87$\% whereas the\\ RF algorithm achieved\\ an accuracy of $99.95$\%\end{tabular}         & \begin{tabular}[c]{@{}l@{}}The model does not\\ classify the type of\\attack taking place and\\work on decreasing the\\latency is needed\end{tabular}                                        \\ \hline 
\cite{obs}                                                                           & \begin{tabular}[c]{@{}l@{}}DoS\\attack,\\ DDoS\\ attacks\end{tabular}                            & \begin{tabular}[c]{@{}l@{}}The data received made\\ obscure by adding some\\ noise and CNN is used\\ to reconstruct the\\ obscured image by\\ using different weights\end{tabular}                                          & \begin{tabular}[c]{@{}l@{}}CNN algorithm is used to \\ reconstruct the obscured\\ data by using some\\ random weights, hence\\ making the data secure.\end{tabular}                                & \begin{tabular}[c]{@{}l@{}}The model re-\\constructed the\\obscured data with\\an accuracy of $81.3$\%\\and it can run on R Pi\\ $3$B as well\end{tabular} & \begin{tabular}[c]{@{}l@{}}Research is open for\\ working on increasing\\the accuracy and\\the efficiency of\\the model\end{tabular} \\ \hline \rowcolor[gray]{0.9}
\cite{GPSD}                                                                          & \begin{tabular}[c]{@{}l@{}}GPS\\ spoofing,\\ DoS\\attack\end{tabular}          & \begin{tabular}[c]{@{}l@{}}The target drone and the\\ size of the drone is\\ detected using bounding\\ box object detection\\ algorithm\end{tabular}                                                                        & \begin{tabular}[c]{@{}l@{}}The bounding box object\\ detection algorithm and the \\ YOLO detection algorithm\\ is used for the real-time\\ detection of the drone\end{tabular}                     & \begin{tabular}[c]{@{}l@{}}This model achieved\\ $77$\% accuracy in\\ detecting the target \\ drone with an average\\ frame rate of $5.22$ fps\end{tabular}        & \begin{tabular}[c]{@{}l@{}}The hunter drone\\is inefficient because\\of its heavy weight\end{tabular}                                           \\ \hline 
\cite{mltab1}                                                                        & \begin{tabular}[c]{@{}l@{}}Jamming,\\ Black\\hole\\attacks\end{tabular} & \begin{tabular}[c]{@{}l@{}}Whenever any event is\\ detected by the UAV, the\\ information is sent to the\\ controller and IDS\\ identifies the malicious\\ node\end{tabular}                                                & \begin{tabular}[c]{@{}l@{}}A hierarchical intrusion\\ detection system is used\\ for detecting the malicious\\ nodes that are injecting\\ false data\end{tabular}                                  & \begin{tabular}[c]{@{}l@{}}The model achieved \\ a detection rate of\\more than $93$\% and a\\false positive rate of\\less than $3$\%\end{tabular}              & \begin{tabular}[c]{@{}l@{}}Implementing the \\model on the swarm\\of drones is a much\\needed work\end{tabular}                                                                                 \\ \hline \rowcolor[gray]{0.9}
\cite{mltab2}                                                                        & \begin{tabular}[c]{@{}l@{}}GPS\\ spoofing\end{tabular}                                          & \begin{tabular}[c]{@{}l@{}}A machine learning-based\\ naive Bayes algorithm is\\ used to check for the\\ presence of the UAV\\ and the classification is\\ done using the k-nearest\\ neighbor algorithm\end{tabular} & \begin{tabular}[c]{@{}l@{}}The naive Bayes algorithm\\ is used for the detection of\\ the micro-UAV and for the\\ classification of the micro-\\ UAV kNN and is used\end{tabular}                  & \begin{tabular}[c]{@{}l@{}}The confusion matrix\\ obtained for the kNN \\ classifier in the model\\ achieved and accuracy\\ of $97.1$\%\end{tabular}               & \begin{tabular}[c]{@{}l@{}}This model can make\\use of a $3$D feature\\cluster map that would\\help improve the real-\\time classification\end{tabular}                                      \\ \hline
\cite{mltab3}                                                                        & \begin{tabular}[c]{@{}l@{}}Jamming\\ attacks,\\ DoS\\attack\end{tabular}                         & \begin{tabular}[c]{@{}l@{}}UAVs are used for data \\ transmission and intrusion\\ detection system is used\\ for detecting any anomaly\end{tabular}                                                                 & \begin{tabular}[c]{@{}l@{}}An intrusion detection\\ system is used for the\\ detection of the \\ anomaly in the network\end{tabular}                                                               & \begin{tabular}[c]{@{}l@{}}An efficient way\\of securing the\\ multi-level ad hoc\\ networks is presented\end{tabular}                                     & \begin{tabular}[c]{@{}l@{}}Other networking \\ solutions can be used\\ to make the model\\more efficient\end{tabular}                                                             \\ \hline \rowcolor[gray]{0.9}
\cite{mltab4}                                                                        & \begin{tabular}[c]{@{}l@{}}DoS\\attack,\\ DDoS\\ attacks\end{tabular}                            & \begin{tabular}[c]{@{}l@{}}Devices selected in white\\ list using the algorithm\\ are only used for\\ data transmission\end{tabular}                                                                        & \begin{tabular}[c]{@{}l@{}}Random Forest algorithm\\ is used for classifying the\\ connected devices as\\ legitimate devices or\\ malicious devices\end{tabular}                                   & \begin{tabular}[c]{@{}l@{}}The model showed an\\ accuracy of  $99.49$\%\\ in detecting the un-\\ authorized device in\\the network\end{tabular}                  & \begin{tabular}[c]{@{}l@{}}The efficient detection\\of a variety of\\compromised drones\\can be worked\\upon in the future\end{tabular}                                                  \\ \hline
\end{tabular}
\label{relml}
\end{table*}
\subsection{\textit{\textbf{\textcolor{black}{Privacy Leakage}}}}
In the case of drones, the authentication algorithms used to enable access to the network are generally cryptographically placed. However, recently, the use of machine learning algorithms to avoid privacy leakage in IoD networks is being explored \cite{IoD}. For avoiding privacy leakage in drones, a pioneer model is proposed in \cite{obs}. Different from \cite{neural}, the authors of \cite{obs} focus on using deep learning algorithms to proactively secure the data, rather than detecting attacks on the network. The authors use deep learning auto-encoders to secure sensor data and to store media on a server. Each bit of data collected from the sensors of the drone is first converted into a digit image of size $28$ by $28$ pixels. Further, some noise is added to the sensor data to make it obscure and it is then sent to a remote cloud server to be saved. Convolutional Neural Network (CNN) is implemented in the reconstruction and classification components. Reconstruction component is used for reconstructing the obscured data into the original data by training the model weights. As the model weights are not known, it becomes almost impossible for the adversary to retrieve the original data. Next, the classification component recognizes the data from the reconstructed data and the digital data is further converted to sensor data by using deep learning auto-encoder. The proposed model is such a lightweight model that it can be run on Raspberry Pi $3$B too. The model was tested on the MNIST dataset, and the results demonstrate an accuracy of $81.3$\% in identifying the reconstructed data. In general, data privacy is ensured by encrypting the data with a variety of cryptographic representations. However, further techniques are required as the privacy techniques using cryptographic keys can be broken once the key is obtained. Another technique that helps in preventing data leakage is Homomorphic authenticated encryption (HAE) \cite{HAE}. Unlike \cite{obs}, the \cite{HAE} model works without the use of a key. HAE allows the users who do not have a key to perform computation on the ciphertext. The computed ciphertext decrypts to the correct function value.

\subsection{\textit{\textbf{\textcolor{black}{Adversarial attacks}}}}
Adversarial machine learning is mainly used to cause a malfunction in a machine learning model by supplying deceptive inputs. The IoD brings in a vast range of sensors, mobile network security issues, and privacy-protecting challenges that are different from traditional internet systems. A model proposed for avoiding adversarial attacks has been proposed in \cite{CNNADV} that uses CNN and RNN for adversary detection. Similar to \cite{RNN,CNN}, the authors of \cite{CNNADV} use the RNN and CNN-based models. However, these models are used to detect and prevent adversarial attacks rather than detecting the presence of drones, as done in  \cite{RNN,CNN}. A pioneer work in detecting and preventing adversarial attacks in IoD is \cite{GPSD}. The authors of \cite{GPSD} use a black and white version of the Tiny You Only Live Once (YOLO) detection system and visual-serving without motion capturing systems. The proposed techniques are even efficient in a GPS denied environment. The proposed model is presented using a drone hunting platform that self localizes using visual inertial odometry (VIO) through ZED stereo camera, and runs a visual tracking and identifying algorithm on Jetson TX$2$. The commands are sent by the algorithm to the PX-$4$ based flight controller. The simulation results demonstrate that the platform could effectively track and chase the adversary. The model achieved $77$\% accuracy with an average frame rate of $5.22$ fps. The proposed work runs significantly faster than other deep learning detection models as mentioned in \cite{CNNADV} and \cite{GIS} with comparable accuracy. Also, it works to detect the adversary in a GPS denied environment, which is not done in other previous works in this direction. However, the fundamental drawback in the proposed model is that the detection algorithm is sensitive to poor lightning.

\textbf{       Summary:} 
\textcolor{black}{The objective here is to enhance the possibilities of adversary drone detection using various machine learning and deep learning approaches. Apart from drone detection, various works have also focused on using such algorithms to prevent attacks in drone communication network. A summary of advantages and disadvantages of major works is shown in Table \ref{mladv}, and a summary of the related works is given in Table \ref{relml}. As seen, ML algorithms have a high capability in detecting unwanted drones and preventing the drones from entering restricted areas. These algorithms are also being widely proposed for secure traffic management and prevention of mid-air collisions.} 

\section{Applications of Fog Computing for Drone Communication Security}

\label{sec8} Fog computing is a powerful complement to cloud computing which can provide a better QoS and can also help in decreasing the security issues in the cloud computing system. It is difficult to connect such a large number of drones directly to the cloud due to high latency delays and unpredictable network connections. Connections between the drones and the fog layer can be easily established with low latency issues. The most important benefit of fog computing is that it does all the computations and keeps the data near to the drone, which keeps it more safe and secure. Fog computing also supports mobility, scalability, heterogeneity as well as platform-independence. \textcolor{black}{Concept of edge computing comes close to fog computing and is said to overlap to a great extent \cite{fog_edge_iot}. Edge computing is for moving the resources from the cloud towards the edge of the network, and is more focused towards the 'things' side. However, fog computing concerns itself mainly with the infrastructure.} In this section, we first of all discuss the basic issues with the traditional approaches that do not use fog computing and then we move on to the challenges in drone communication and possible fog computing based solutions. 
 
There are various traditional methods that help in securing drone communication without leveraging the benefits of fog computing. A traditional man-in-the-middle attack detection system has been proposed in \cite{non5}, which uses the precise timing of arrival of data packets to infer the possibility of the attack. If the packet arrives late than the expected threshold time, the possibility of the attack is inferred. This method can fail in several circumstances where heavy background noise is present as the arrival of the data packet highly depends on the transmission channel. Bamasag et al. in \cite{non6}, proposed a multicast authentication model for data transmission in the desired time interval. This model makes the use of Shamir’s secret sharing technique \cite{non7} in which the secret can be unlocked if the authenticator has enough number of shares. Although this method provides some reliability, storing such a large number of keys is not preferred considering the resource constrained nature of drones.

We  further  discuss  the  list  of  specific  security  issues  that can be resolved and prevented using fog-computing as a solution.
 \begin{table*}[]
\centering
\caption{A summary of advantages and disadvantages of major applications of Fog Computing for drone communication security.}
\begin{tabular}{|l|l|l|l|}
\hline \rowcolor[gray]{0.8}
\begin{tabular}[c]{@{}l@{}}Major\\ Approaches\end{tabular} & Advantages & Disadvantages  & \begin{tabular}[c]{@{}l@{}} \textcolor{black}{Benefits over}\\ \textcolor{black}{traditional approaches} \end{tabular}\\
\hline
\cite{27}                                 & \begin{tabular}[c]{@{}l@{}}• High performance and accuracy as the\\ spoofing attack can be detected from\\ $10$ meters and in $250$ milli-seconds\end{tabular}                                & \begin{tabular}[c]{@{}l@{}}• Cannot efficiently avoid collisions\\ and also sometimes fails in\\ detecting the obstacles\end{tabular}   & \begin{tabular}[c]{@{}l@{}} \textcolor{black}{Easy deployment,}\\ \textcolor{black}{Confidentiality} \end{tabular}                  \\
\hline
\cite{IDS}                                & \begin{tabular}[c]{@{}l@{}}• A very high efficiency and a low\\ resource-demanding model\end{tabular}               & \begin{tabular}[c]{@{}l@{}}• Highly dependent on network's\\ latency which demands more\\ optimization\end{tabular}  &  \begin{tabular}[c]{@{}l@{}}\textcolor{black}{Security against man-}\\ \textcolor{black}{in-the-middle attacks}  \end{tabular} \\
\hline
\cite{FSS}                                & \begin{tabular}[c]{@{}l@{}}• Identity-based authentication enhances\\ end-to-end security between the edge\\ layer and the fog layer\end{tabular}                                                 & \begin{tabular}[c]{@{}l@{}}• The data processing time highly\\ varies with the configuration of the\\ device used for detection\end{tabular} & \begin{tabular}[c]{@{}l@{}} \textcolor{black}{Authentication, Data}\\ \textcolor{black}{integrity, Non-repudiation} \end{tabular} \\
\hline
\cite{fogres}                             & \begin{tabular}[c]{@{}l@{}}• The architecture covers all the three\\ aspects i.e. minimizing the latency and\\ the energy consumption and maximizing\\ the reliability in the drones\end{tabular} & \begin{tabular}[c]{@{}l@{}}• The architecture is not very fast\\ and future work is needed for\\ increasing the efficiency of the\\ architecture\end{tabular} & \begin{tabular}[c]{@{}l@{}} \textcolor{black}{Better performance than} \\ \textcolor{black}{ LRGA-MIE and LP-based} \\ \textcolor{black}{algorithms} \end{tabular}\\
\hline
\cite{cache}                              & \begin{tabular}[c]{@{}l@{}}• The model decreases the latency\\ experienced by the drone and increases\\ the Quality-of-Experience (QoE)\end{tabular}
&\begin{tabular}[c]{@{}l@{}} • Highly sensitive to the number\\ of drones \end{tabular} & \begin{tabular}[c]{@{}l@{}} \textcolor{black}{Low latency, High} \\ \textcolor{black}{QoE} \end{tabular}\\  \hline      
 \end{tabular}
\label{fogadv}
\end{table*}

\subsection{ \textit{\textbf{\textcolor{black}{GPS spoofing attacks}}}}
UAVs in the fog environment are susceptible to a lot of challenges against its benefits like mobility, scalability, and accurate location tracking. GPS spoofing is a notable security breach attack that sends incorrect GPS information to the receiver. UAVs need special attention since traditional internet systems like cloud computing in the former causes latency overloads and unforeseeable network issues. In the past, there are various GPS spoofing detection methods that have been adopted. The major ones are detection based on cryptographic algorithms and using auxiliary equipment as mentioned in \cite{27}. The authors take flight security and safety of drones, acting as fog nodes in an airborne fog computing system, in consideration. The model uses visual sensors combined with IMU (Inertial Measurement Unit) for information fusion to solve GPS spoofing issues. Using DJI Phantom $4$ with a frame rate of $30$ fps, it was observed that the spoofing attack can be detected from $10$ meters and in $250$ milli-seconds. The authors of \cite{fogt1} propose a fog-to-cloud computing framework for a Dragnet based amateur drone surveillance system. GPS spoofing and jamming attacks can be detected using the framework, which is inspired from traditional military anti-drone technologies. The amateur surveillance system is empowered with a "brain" for high-level intelligence consisting of a fog-to-cloud model. It is a system of coordinated measures for sensing a spoofing attack on the system by global decision-making based on the actions on the amateur drones. The Kashmar incident as mentioned in \cite{26} could've been prevented if US had employed some of the frameworks of fog computing as mentioned above.

\subsection{ \textit{\textbf{\textcolor{black}{Man-In-The-Middle attack}}}}
Man-in-the-middle attack problem threatens the very demanding aspect of IoD, which is integrity, as mentioned in \cite{int}. A pioneer work in the direction of low resource demanding model with a high level of security to prevent man-in-the-middle attack in IoD is proposed in \cite{IDS}. The authors propose an intrusion detection system (IDS) and intrusion prevention system (IPS) for preventing man-in-the-middle attack at the fog layer. Although, the model is proposed for IoT devices in general, it can be easily implemented on IoD as well. In the proposed network model, IDS nodes are deployed at a one-hop distance. Whenever an IDS node finds a compromised node or an intruder, it simply indicates the nodes in its proximity to cut off connection with the compromised node. On deployment, IDS nodes acquire the key from the cloud and distribute it to fog nodes. To prevent intrusion, all the packets are encrypted using Advanced Encryption System (AES), and Diffie-Hellman key exchange \cite{AES} is used for key exchange. IDS nodes periodically interrogate the fog nodes and observe the receiver’s behavior. IDS nodes expect the receiver to decrypt the packet in some pre-defined time. If the round trip time of interrogation exceeds the pre-defined time, the IDS concludes that the fog node is malicious. Additionally, if an attacker knows the existence of IDS nodes and the protocol they are using, he still does not know the nature of interrogation which is pre-programmed before the deployment of the nodes. This further reduces the chances of the attack. The proposed model, when implemented at the fog layer, could help in the identification and prevention of the man-in-the-middle attacks so that manipulated information does not reach the cloud. The simulation of the model was done over OMNET++. Latency overhead for deploying IDS and IPS was $40$ milli-seconds. Time taken to detect an attack was discovered to be between $2.48$ seconds to $2.53$ seconds. Since $2$ seconds was the time between investigation sessions, actual discovery time was approximately $0.5$ seconds. Energy overhead incurred on the fog nodes by IDS nodes was negligible which makes it a very less resource-demanding model for detecting man-in-the-middle attack. However, in the proposed model, the investigating-time is inversely proportional to the network's latency and energy overhead of the IDS network model. Further work is required in the direction of optimization algorithms to improve the efficiency of the proposed model.
 
 \begin{table*}[]
\centering
\caption{Applications of Fog computing for drone communication security.}
\begin{tabular}{|p{0.033\linewidth}|p{0.11\linewidth}|p{0.235\linewidth}|p{0.114\linewidth}|p{0.189\linewidth}|p{0.17\linewidth}|}
\hline \rowcolor[gray]{0.7}

Ref.                           & Security Issues                                                                            & Mechanism                                                                                                                                                                                                                                    & \begin{tabular}[c]{@{}l@{}}Fog Computing\\ Feature used\end{tabular}                                                & Major achievement                                                                                                                                                                                          & Open issues                                                                                                                                                \\
\hline
\cite{27}     & \begin{tabular}[c]{@{}l@{}}GPS spoofing,\\ Eavesdropping\end{tabular}                              & \begin{tabular}[c]{@{}l@{}}The visual sensors combined\\ with IMU are deployed on\\ the drone and the data is\\ transmitted to the fog layer\\ detecting the attack\end{tabular}                                                             & \begin{tabular}[c]{@{}l@{}}Distributed\\ computing,\\ Scalability\end{tabular}                                      & \begin{tabular}[c]{@{}l@{}}The spoofing attack can\\ be detected from $10$\\ meter far and in $250$\\ milli-seconds.\end{tabular}                                                                          & \begin{tabular}[c]{@{}l@{}}Methods for avoiding \\ mid-air collisions\\can also be \\implemented\end{tabular}                                               \\
\hline \rowcolor[gray]{0.9}
\cite{IDS}    & \begin{tabular}[c]{@{}l@{}}Man-in-the-\\ middle attacks\end{tabular}                               & \begin{tabular}[c]{@{}l@{}}The malicious data is\\ prevented from entering the\\ cloud layer as the attack is\\ detected in the fog layer by\\ using the cryptographic keys\end{tabular}                                                     & \begin{tabular}[c]{@{}l@{}}High\\computation\\ power\end{tabular}                                                    & \begin{tabular}[c]{@{}l@{}}The model is very less\\ resource-demanding\\ and can detect the attack\\ in less than $0.5$ secs\end{tabular}                                                                  & \begin{tabular}[c]{@{}l@{}}Future work on\\increasing the\\efficiency of the\\model is much\\needed\end{tabular}   \\
\hline
\cite{FSS}    & \begin{tabular}[c]{@{}l@{}}Eavesdropping,\\ Man-in-the-\\ middle attacks,\\ Hijacking\end{tabular} & \begin{tabular}[c]{@{}l@{}}The drone gets authenticated\\ from the fog layer that\\ contains the hashing\\ algorithm, and then only is\\ allowed to transmit data\end{tabular}                                                               & \begin{tabular}[c]{@{}l@{}}Low latency,\\ Mobility,\\ Heterogeneity\end{tabular}                                    & \begin{tabular}[c]{@{}l@{}}The average end-to-end\\ processing time was\\ $2.59$ secs and the\\ average overall response\\ time was $3.17$ secs\end{tabular}                                               & \begin{tabular}[c]{@{}l@{}}The average overall\\ response time highly\\ varied with the\\number of devices\end{tabular}                                   \\
\hline \rowcolor[gray]{0.9}
\cite{fogres} & \begin{tabular}[c]{@{}l@{}}Latency,\\ Resource\\ constraints\end{tabular}                          & \begin{tabular}[c]{@{}l@{}}The task is divided into\\ several small task using\\ ADMM algorithm and is\\ transmitted to the nearby\\ ready drones that completes\\ the task and transmit back\\ the result acting as a fog node\end{tabular} & \begin{tabular}[c]{@{}l@{}}Low latency,\\ High\\ computation\\ power\end{tabular}                                     & \begin{tabular}[c]{@{}l@{}}The model showed\\ positive results in\\ minimizing the latency\\ and the energy\\ consumption and\\ maximizing the\\ reliability in the drones\end{tabular}                    & \begin{tabular}[c]{@{}l@{}}Future work is \\required in making\\the model more\\ reliable and safe\\ for the drones\end{tabular}                            \\
\hline
\cite{fogt1}  & GPS spoofing                                                                                       & \begin{tabular}[c]{@{}l@{}}The surveillance devices\\ acting as a fog layer sends\\ the data to the cloud layer\\ and gets the result back and\\ transmits to the amateur\\ drone for implementation\end{tabular}                            & \begin{tabular}[c]{@{}l@{}}High\\computation\\ power,\\ Distributed\\ computing\end{tabular}                         & \begin{tabular}[c]{@{}l@{}}The model detects the\\ authorized drone with a\\ greater probability than\\ detecting the false or\\ unauthorized drone\end{tabular}                                           & \begin{tabular}[c]{@{}l@{}}To efficiently detect \\the drone, a high\\detection delay is\\expected from the\\ model which should\\ be reduced\end{tabular} \\
\hline \rowcolor[gray]{0.9}
\cite{fogt2}  & \begin{tabular}[c]{@{}l@{}}DoS attacks,\\ DDoS attacks\end{tabular}                                & \begin{tabular}[c]{@{}l@{}}The serial number of each\\ device is stored in the fog\\ layer and whenever any\\ device wants to communicate\\ with the other device it needs\\ to verify its serial number\\ with the fog layer\end{tabular}   & \begin{tabular}[c]{@{}l@{}}Low latency,\\ High \\computation\\ power \end{tabular} & \begin{tabular}[c]{@{}l@{}}The model uses very \\less bandwidth and\\increases the security\\ in the IoT devices\\as no device can\\communicate without\\authentication\end{tabular} & \begin{tabular}[c]{@{}l@{}}The model can be\\ implemented on the\\drones specifically to\\increase their security\end{tabular}   \\   \hline   \rowcolor[gray]{0.9}             
                     
\end{tabular}
\label{relfog}
\end{table*}

\subsection{ \textit{\textbf{\textcolor{black}{Eavesdropping}}}}
Eavesdropping is an attack that affects the confidentiality of the data in the drones \cite{int}. Classical security solutions exist such as Secure Sockets Layer (SSL) \cite{SSL}, but it cannot be implemented on drones as they lack enough memory and CPU power to perform the required cryptographic operations. Therefore, offloading the additional security operations to a more resourceful entity such as fog nodes is a promising solution. A model that addresses this problem in drones is proposed in \cite{FSS}. Fog Security Service (FSS) mechanism is proposed that uses public and private key cryptography schemes. It consists of a Verifier, PKG (Private key generator), and a hashing algorithm at the fog layer. In the proposed model, input security parameters that include an identifier (unique), username, and password for verification of the sender are assigned to every drone. PKG is used for communication between the fog layer and the edge layer. After node authentication, asymmetric encryption is used for getting symmetric keys from the fog layer. For public-key encryption, the Rivest-Shamir-Adleman (RSA) algorithm \cite{RSA} is used. Nonce values are also used for preventing play-back attacks. FSS provides identity-based authentication as the private key is used for encryption and decryption both. This enhances the end-to-end security between the edge and the fog layer. For IoD networks, ground access points along with UAVs are present \cite{fogaccess}. Therefore, installing the proposed FSS layer along the data transmission paths could identify and prevent eavesdropping problems in the drones. OPNET based network simulator is used to evaluate the proposed method. In addition to different traffic loads, several devices representing different capacities and resources were used for experimentation. The performance of the model was evaluated based on the response time. The average E2E processing time was $2.59$ secs, while the overall response time was $3.17$ secs on average. Response time was measured against the state-of-the-art Authentication Proxy as a Service (ApaaS) \cite{ApaaS} and legacy methods. Processing time varied according to different hardware used. However, the heterogeneity of drones created a lot of dependencies related to processing time. Decreasing the variance of time involved based on heterogeneity is open for research.

\subsection{\textit{\textbf{\textcolor{black}{Resource constraint issues}}}}
As discussed above, drones have several applications like delivering products, military applications, etc. Therefore, drones need to have high computation power. The latency-sensitive applications such as disaster management, path recognition are also at risk due to this issue. A Fog Computing aided Swarm of Drones (FCSD) architecture is proposed in \cite{fogres}, which helps in minimizing the latency in drone communication. As the drones are highly resource-constrained, the task is divided into several small tasks using a Proximal Jacobi Alternating Direction Method of Multipliers (ADMM) \cite{ADMM}. ADMM is an algorithm that distributes a task into several small tasks and assigns it to the devices connected in the network that are ready to perform the task. An initiator drone is used to assign the tasks to the nearby drones using the ADMM algorithm. The drones complete the specified task and transmit the computed results back to the initiator drone. The simulation results demonstrate a considerable improvement in terms of reduction in transmission latency and computation latency. The energy consumption including transmission energy consumption and the computation energy consumption is also considered for the FCSD to reduce the overall energy consumption in the drone. The ADMM algorithm results in better performance when compared with the baseline pre-existing algorithms such as the latency and reliability constrained minimum energy consumption algorithm based on genetic algorithm (LRGA-MIE) \cite{LRGA} and a newly developed Linear Programming (LP) based algorithm. The Proximal Jacobi ADMM based algorithm gave the optimal solution and algorithm converged after the $14^{th}$ iteration. Another model for minimizing the latency in the swarm of drones that uses a decentralized algorithm for task allocation based on game theory is discussed in \cite{ano}. However, this model fails in providing the level of reliability provided by the ADMM algorithm. Moreover, the model in \cite{ano} converges towards the optimal solution after a large number of iterations as compared to \cite{fogres}.

\subsection{\textit{\textbf{\textcolor{black}{Minimum Latency in data dissemination}}}}
The dissemination of the data is required to have the least possible latency and fallacy. A model known as edge caching is proposed in \cite{cache} in which common files that the drone captures are cached and are made available whenever needed. The data that the user demands is generated by merging the data files collected by the different sensors installed in the UAV. The common data files can be stored in the cache-enabled UAV, which will then be transmitted directly to the demanding user. This model helps in decreasing the latency and increasing the Quality-of-Experience (QoE) as it has the common data already cached, which collectively helps in generating the demanded data. However, this model suffers from the drawback that whenever the number of drones increases, the data transmitting power decreases. The simulation results demonstrate that the transmission power decreases by $86$\% when the number of UAVs is increased from $3$ to $7$. Hence this model is highly sensitive to the number of UAVs. However, another similar model proposed in \cite{inter} gives a significant performance as compared to the mechanism proposed in \cite{cache}. The authors of \cite{inter} use some ML algorithms such as CNN and RNN for classifying the already existing data, and the required data for the generation of the demanded data. The use of these algorithms significantly improves the system of the model even with the increased number of drones.

\textbf{       Summary:}
\textcolor{black}{This section portrays that fog computing can help in preventing various attacks like GPS spoofing, man-in-the-middle attacks, and eavesdropping attacks in drone communication. Fog computing majorly works in minimizing the latency in drone communication considering the resource-constrained nature of drones. A summary of the advantages and disadvantages of major works that use fog computing as a solution to drone communication security are presented in Table \ref{fogadv}, and a summary of all the related works in this direction is presented in Table \ref{relfog}. As seen, fog computing minimizes the load on the cloud and helps the drone to offload its tasks to the fog layer, thereby minimizing the latency and maximizing the reliability in drone communication.}

\section{\textcolor{black}{Lessons Learned,} Future Research Directions and Open Challenges}
\label{sec9}
\subsection{\textcolor{black}{Lessons Learned}}
The UAV industry is growing rapidly, and as the applications come to use, we face various challenges that still need to be handled. Although various technologies mentioned above are anticipated to help secure drone communications, there are various constraints in these technologies as well. It is required to closely focus on the solution constraints as well before implementing these solutions in different drone applications.

Blockchain technology is itself an emerging technology and has not been well implemented and tested in non-financial domains.
\textcolor{black}{Blockchain technology has properties that can help secure drone communication across application areas like mining, delivery, surveillance, and disaster management effectively because it improves data security (against DoS, jamming, GPS spoofing, eavesdropping and wormhole attacks) and transparency, even in a swarm of drones.}

\textcolor{black}{SDN aims at making the network more agile, flexible, and secure (against DoS, jamming, GPS spoofing and black hole attacks) because of its infrastructure of separating the data plane and the control plane. These incentives make drone communication networks useful in application areas of military, photography, and $5$G networks.}

\textcolor{black}{Use of different approaches of machine learning algorithms depend on the application area and domain. ML can be used for securing drone communication networks (against DoS, GPS spoofing, jamming, wormhole attacks), as well as physical security (drone detection) of the drones. Characteristics of frameworks using ML make it suitable for drone applications like traffic management, fault detection, and navigation systems.}

\textcolor{black}{Fog computing provides a better QoS, scalability, flexibility, low latency, platform-independence, and improves the security of the network (against GPS spoofing, man-in-the-middle, eavesdropping, hijacking, and DoS attacks. All these advantages make such a framework suitable for drone application areas involving big data, smart vehicles, and energy conservation.}

\subsection{\textcolor{black}{Future Research Directions and Open Challenges}}
Some of the future research directions in this field are as follows:
\begin{itemize}
\item The drones are resource-constrained devices. Implementing security algorithms such as blockchain, in a swarm of drones, means adding some more storage capacity and computation ability on drones. This might end up reducing the flight time. Moreover, using blockchain for non-critical communication can be acceptable, but for the critical things like location coordinates, blockchains can cause high latency as of now. Further research is required to enable the security algorithms keeping the resource-constrained nature of drones in mind.

\item The gateways between drones, ground controllers and satellites in drone communication are also highly vulnerable to various security attacks. If the gateways are compromised, then the whole network is compromised, even though the end devices are highly secure. Further analysis is required on how to secure the gateways between different hops in drone communication.

\item The current architecture of fog computing does not support the inter-fog resource and task sharing. In drone communication, few fog nodes or access points might be less loaded as compared to others. In such scenarios, the fog nodes can directly interact with each other and can share the tasks among themselves. This could further reduce the transfer of data from fog to cloud, thereby enhancing security. 

\item The current blockchain architecture is highly limited in terms of the number of nodes in permission-ed networks and in terms of throughput in permission-less networks. Various consensus algorithms are being designed to support high throughput along with a large number of nodes or users. 

\item A concept of multi and distributed controllers is being proposed in some works to overcome the problem of the controller being a single point of failure in SDN architectures. However, further work is required to ensure secure and near real-time communication between different controllers in SDN. 



\end{itemize}

\section{Conclusion}
\label{sec10}In this study, \textcolor{black}{we gave an overview of the various application areas of drones. Following which, security threats (drone-specific and IoT-generic) and potential vulnerabilities in specific applications of drone communications were explained. Furthermore, a brief overview of the fundamentals for various technologies is given.} Existing and upcoming solutions to overcome the security threats in drone applications \textcolor{black}{using different concepts have been discussed in detail in the subsequent sections}. The major technologies covered in solution architectures include software-defined networks, blockchain, fog/edge computing, and machine learning. Detailed benefits of these technologies to overcome the security threats in specific drone applications have also been discussed. The state-of-the-art drone security has also been discussed with some improvement suggestions, open issues, and future research directions. This survey is expected to serve as a valuable resource for security enhancement for upcoming and existing drone applications. 
\bibliography{main.bib}
\bibliographystyle{IEEEtran}

\vskip -1\baselineskip plus -1fil
\begin{IEEEbiography}[{\includegraphics[width=1in,height=1.25in,clip,keepaspectratio]{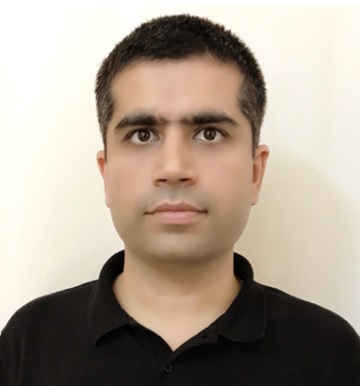}}]{Vikas Hassija} received the B.Tech. degree from Maharshi Dayanand University, Rohtak, India, in 2010, and the M.S. degree in telecommunications and software engineering from the Birla Institute of Technology and Science (BITS), Pilani, India, in 2014. He is currently pursuing the Ph.D. degree in IoT security and blockchain with the Jaypee Institute of Information and Technology (JIIT), Noida, where he is currently an Assistant Professor. His research interests include the IoT security, network security, blockchain, and distributed computing.
\end{IEEEbiography}

\vskip -1\baselineskip plus -1fil
\begin{IEEEbiography}[{\includegraphics[width=1in,height=1.25in,clip,keepaspectratio]{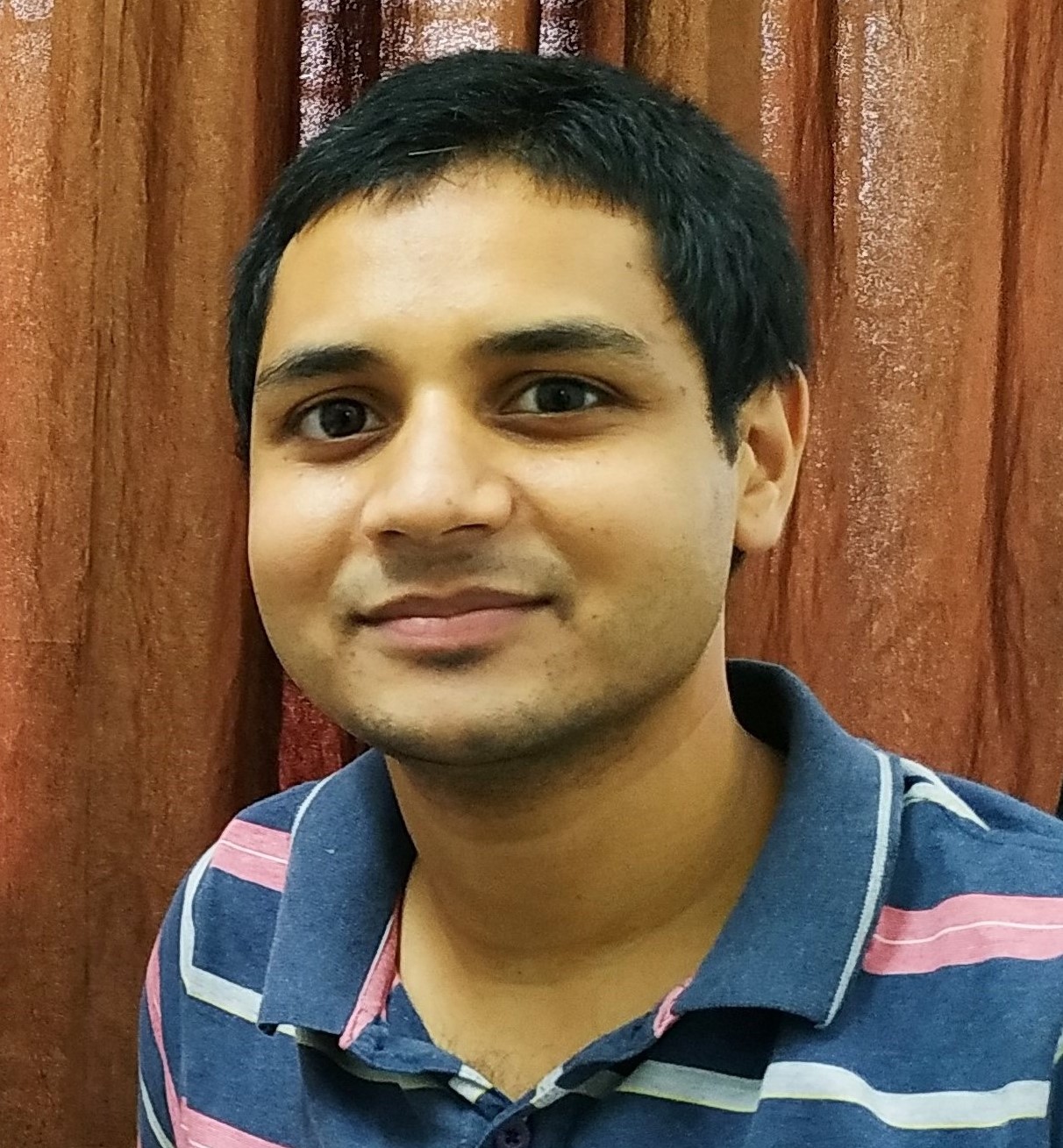}}]{Vinay Chamola}\textnormal{ received the B.E. degree in electrical and electronics engineering and master’s degree in communication engineering from the Birla Institute of Technology and Science, Pilani, India, in 2010 and 2013, respectively. He received his Ph.D. degree in electrical and computer engineering from the National University of Singapore, Singapore, in 2016. He is currently Assistant Professor with the Department of Electrical and Electronics Engineering, BITS-Pilani, Pilani where he heads the Internet of Things Research Group / Lab. He has over 62 publications in high ranked SCI Journals including more than 46 \textit{IEEE Transaction} and Journal articles. His research interests include IoT Security, Blockchain, UAVs, VANETs, 5G and Healthcare. He is an Associate editor of the IEEE Internet of Things Magazine, IET Networks and IET Quantum Communications Journal. He also serves as an Area Editor of the Adhoc networks journal, Elsevier. He is a senior member of the IEEE. 
}
\end{IEEEbiography}

\vskip -2\baselineskip plus -1fil
\begin{IEEEbiography}[{\includegraphics[width=1in,height=1.5in,clip,keepaspectratio]{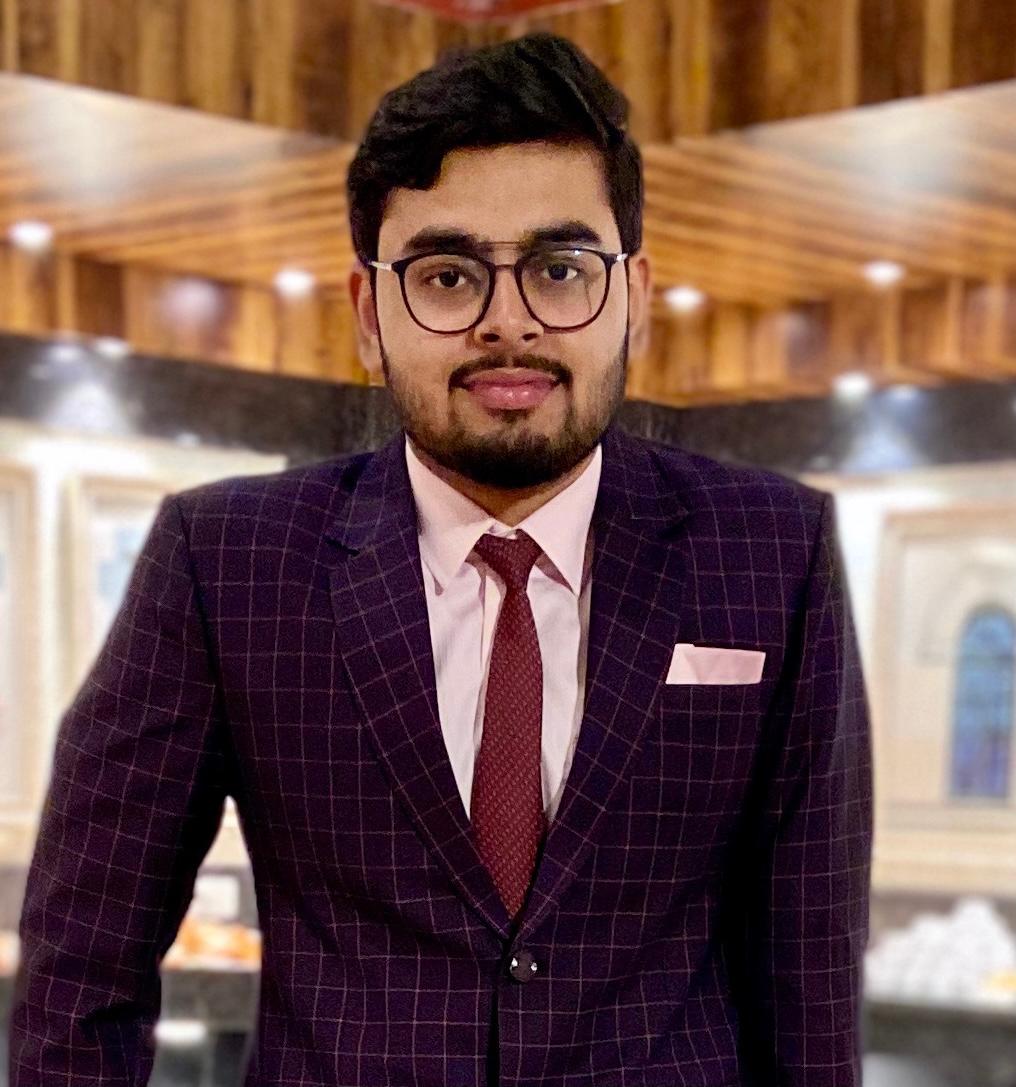}}]{Adhar Agrawal} is currently pursuing a B.Tech. degree from the Jaypee Institute of Information Technology (JIIT), Noida. He has completed a few projects in the field of evolutionary algorithms, machine learning and data analytics. He was pursuing his research internship at the Birla Institute of Technology and Science (BITS), Pilani under Dr. Vinay Chamola. His research interests include distributed ledger technology, evolutionary computing, machine learning, and fog computing.
\end{IEEEbiography}

\vskip -2\baselineskip plus -1fil
\begin{IEEEbiography}[{\includegraphics[width=1in,height=1.5in,clip,keepaspectratio]{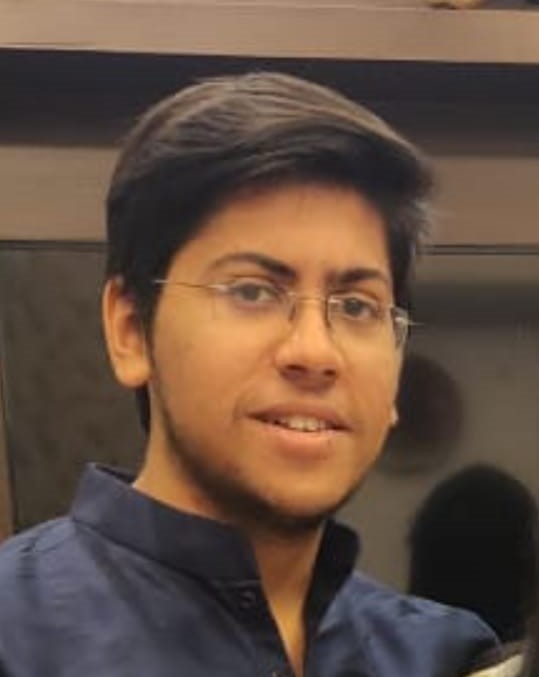}}]{Adit Goyal} is currently pursuing the B.Tech. degree with the Computer Science department, Jaypee Institute of Information Technology (JIIT), Noida. He has completed a few projects in the field of data science, machine learning, and big data. He is currently pursuing his research internship with BITS-Pilani, Pilani, India under Dr. V. Chamola. His research interests include machine learning, data science, and quantum computing.
\end{IEEEbiography}

\vskip -2\baselineskip plus -1fil
\begin{IEEEbiography}[{\includegraphics[width=1in,height=1.5in,clip,keepaspectratio]{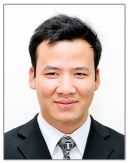}}]{Nguyen Cong Luong} is currently a lecturer in the Faculty of Computer Science, PHENIKAA University, Hanoi, Vietnam. He is also a researcher in the PHENIKAA Research and Technology Institute (PRATI), Hanoi, Vietnam. His research interests include signal processing and resource management in wireless networks.
\end{IEEEbiography}

\vskip -1\baselineskip plus -1fil
\begin{IEEEbiography}[{\includegraphics[width=1in,height=1.25in,clip,keepaspectratio]{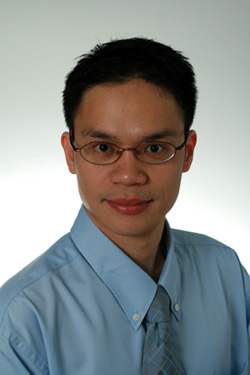}}]{Dusit Niyato} (M’09-SM’15-F’17) is currently a Professor in the School of Computer Science and Engineering, at Nanyang Technological University, Singapore. He received B.Eng. from King Mongkuts Institute of Technology Ladkrabang (KMITL), Thailand in 1999 and Ph.D. in Electrical and Computer Engineering from the University of Manitoba, Canada in 2008. His research interests are in the area of Internet of Things (IoT) and network resource pricing.
\end{IEEEbiography}

\vskip -1\baselineskip plus -1fil
\begin{IEEEbiography}[{\includegraphics[width=1in,height=1.25in,clip,keepaspectratio]{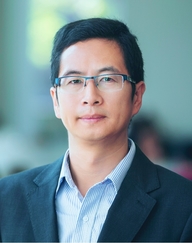}}]{Fei Richard Yu} is a professor at Carleton University, Canada. His research interests include blockchain, security, and green ICT. He serves on the Editorial Boards of several journals, and is a Co-Editor-in-Chief for Ad Hoc \& Sensor Wireless Networks, and is a Lead Series Editor for IEEE Transactions on Vehicular Technology and IEEE Communications Surveys \& Tutorials. He has served as a Technical Program Committee (TPC) Co-Chair of numerous conferences.
\end{IEEEbiography}

\vskip -2\baselineskip plus -1fil
\begin{IEEEbiography}[{\includegraphics[width=1in,height=1.25in,clip]{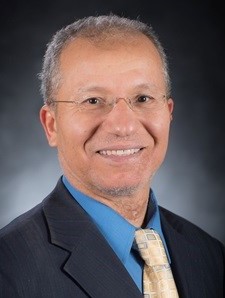}}]{Mohsen Guizani}\textnormal{(S’85–M’89–SM’99–F’09) received the B.S. (with distinction) and M.S. degrees in electrical engineering, the M.S. and Ph.D. degrees in computer engineering from Syracuse University, Syracuse, NY, USA, in 1984, 1986, 1987, and 1990, respectively. He is currently a Professor at the Computer Science and Engineering Department in Qatar University, Qatar. Previously, he served in different academic and administrative positions at the University of Idaho, Western Michigan University, University of West Florida, University of Missouri-Kansas City, University of Colorado-Boulder, and Syracuse University. His research interests include wireless communications and mobile computing, computer networks, mobile cloud computing, security, and smart grid.
}
\end{IEEEbiography}

\end{document}